\documentclass[10pt,journal,compsoc]{IEEEtran}
\usepackage{enumitem}
\usepackage{tikz}
\usepackage{amsmath}
\usepackage{float}
\usepackage{enumitem}

\usepackage{hyperref}
\usepackage{placeins,adjustbox}
\usepackage{threeparttable}
\usepackage{xcolor}
\usepackage{comment,booktabs}
\usepackage{xspace,setspace}
\usepackage{graphicx}
\usepackage{balance}
\usepackage[frozencache,cachedir=.]{minted}
\usepackage[style=trad-plain]{biblatex}
\newcommand{\tool}{{\sc PolyScope}\xspace}
\newcommand*\circled[1]{\tikz[baseline=(char.base)]{
            \node[shape=circle,draw,inner sep=2pt] (char) {#1};}}

\newcommand{\myparagraph}[1]{\vspace{0.25em}\noindent\textbf{#1:}}
\let\paragraph=\myparagraph

\newcommand{\trent}[1]{\textcolor{orange}{\textbf{TRENT:} #1}}
\newcommand{\eddy}[1]{\textcolor{brown}{\textbf{EDDY:} #1}}

\newcommand{\haining}[1]{\textcolor{green}{[\textbf{haining:} #1]}}
\newcommand{\zhiyun}[1]{\textcolor{cyan}{[\textbf{Zhiyun:} #1]}}

\ifCLASSINFOpdf
\else
\fi

\begin{document}

\title{PolyScope: Multi-Policy Access Control Analysis to Triage Android Scoped Storage}
%
%
%
%

\author{Yu-Tsung~Lee,
        Haining~Chen, William~Enck, Hayawardh~Vijayakumar, Ninghui~Li, Zhiyun~Qian, Giuseppe~Petracca~
        and~Trent~Jaeger,~\IEEEmembership{IEEE~Senior Member}
\IEEEcompsocitemizethanks{

\IEEEcompsocthanksitem Yu-Tsung Lee and Trent Jaeger are with Penn State University. \protect\\
E-mail:  yxl74@psu.edu,trj1@psu.edu
\IEEEcompsocthanksitem Haining Chen is with Google. \protect\\
E-mail: hainingc@google.com
\IEEEcompsocthanksitem William Enck is with North Carolina State University.\protect\\
E-mail:whenck@ncsu.edu
\IEEEcompsocthanksitem Hayawardh Vijayakumar is with Samsung Research North America.\protect\\
E-mail:h.vijayakuma@samsung.com
\IEEEcompsocthanksitem Ninghui Li is with Purdue University.\protect\\
E-mail:ninghui@cs.purdue.edu
\IEEEcompsocthanksitem Zhiyun Qian is with UC Riverside.\protect\\
E-mail:zhiyunq@cs.ucr.edu
\IEEEcompsocthanksitem Giuseppe Petracca's work was done when he was a student at Penn State.\protect\\
E-mail:petracca.giuseppe@gmail.com \protect\\}
\thanks{Manuscript received April 19, 2022; revised August 26, 2022.}}

\IEEEtitleabstractindextext{%
\begin{abstract}
Android's filesystem access control is its foundation for system integrity.
It combines mandatory (e.g., SELinux) and discretionary (e.g., Unix permissions) access control with other specialized access controls (e.g., Android permissions), aiming to protect 
Android/OEM services from third-party applications.  However, OEMs often introduce vulnerabilities when they add market-differentiating features because they fail to correctly reconfigure this complex combination of policies. 
In this paper, we propose the \tool tool to triage Android  filesystem access control policies to find the authorized operations that may be exploited by adversaries to escalate their privileges, called {\em attack operations}.  
In this paper, we demonstrate the effectiveness of \tool by assessing the impact of the recently introduced {\em Scoped Storage} defense for Android.  \tool introduces three major advantages over prior access control policy analyses for this analysis: (1)~independent extension and analysis of individual policy models, to ease the addition of Scoped Storage; (2)~knowledge of the flexibility that untrusted parties have to modify access control policies; and (3)~the ability to identify attack operations that system configurations allow.  We apply \tool to three Google and five OEM Android releases, finding that Scoped Storage reduces the number of attack operations possible on external storage resources by over 50\%.  However, we also find two previously unknown vulnerabilities because OEMs only adopt Scoped Storage partially, limiting its benefit. 
Thus, we show how to use \tool to assess an ideal scenario where all apps are compliant to Scoped Storage, which can 
the number of untrusted parties that can access attack operations by over 65\% on OEM systems. 
As a result, we find that \tool can help Android OEMs triage complex access control policies to identify the specific attack operations worthy of further examination.
\end{abstract}

}

\maketitle

\IEEEdisplaynontitleabstractindextext

%
\IEEEpeerreviewmaketitle

\IEEEraisesectionheading{\section{Introduction}\label{sec:introduction}}

\IEEEPARstart{A}{ndroid} has become the most dominant mobile OS platform worldwide, deployed by a large number of vendors across a wide variety of form factors, including phones, tablets, and wearables~\cite{statcounter}.  With Android's increased integration into people's daily lives, Android needs to provide sufficient and appropriate assurances of platform integrity.  Additionally, vendors must be able to extend the Android platform to support their custom functionality and yet maintain such assurances to their customers.  Android's implementation of filesystem access control is one of the most important defenses for providing such assurances.

Despite aggressively adopting advanced mandatory access control (MAC) methods (e.g., Security Enhanced (SE) Android~\cite{smalley2013security})  
in combination with traditional discretionary access control (DAC) and developing specialized access control solutions like {\em Scoped Storage}~\cite{scopedstorage,lee21ieeesp}, 
Android continues to report filesystem vulnerabilities. 

One recent case reported by Checkpoint~\cite{MITD} shows how an untrusted application can abuse write permission to Android's external storage to maliciously replace a victim application's library files before it installs them, which is an example of a {\em file squatting attack}. Such an attack enables adversaries to trick victims into using (e.g., executing, in this example) adversary-controlled files.  Additionally, a report from Google \cite{cve202220220} shows how a vulnerability in the ContactsProvider allowed untrusted applications to open/delete/insert files in unauthorized location by providing a maliciously crafted URI. Furthermore, adversary could potentially lead ContactsProvider to access a symbolic link file which would be an example of {\em link traversal attack}.  This vulnerability could have a serious negative impact, as ContactsProvider is a very privileged process and have access to sensitive files.  Researchers have shown that these two classes of attacks are possible when the access control policy allows an adversary to write filesystem resources used by the victim~\cite{sting}.

To detect these filesystem vulnerabilities, researchers have proposed to apply automated access control policy analysis techniques~\cite{jaeger02sacmat,setools} to Android systems~\cite{enck09ccs,wae+17,Wang2015,aafer15ccs}.  Access control policy analyses transform individual policy rules into the information flows that they authorize to enable detection of secrecy (e.g., data leakage) and integrity (e.g., use of untrusted executables) problems.  Recent work computes the information flows of combined MAC and DAC policies~\cite{chen17acsac}, as well as including Linux capabilities~\cite{BigMAC}.  However, detecting authorized information flows on current access control policies is insufficient to detect attacks in two ways. First, such techniques miss some possible attacks, including the Checkpoint~\cite{MITD} and ContactsProvider~\cite{cve202220220} attacks, because they lack the ability to capture how adversaries may broaden their privilege
by manipulating the inherent flexibility in the Unix and Android permission systems. Second, these techniques may also identify many spurious threats, because they do not determine whether the authorized information flows found can really be exploited to launch attacks given the system configurations.

In this paper, we present a novel access control policy analysis tool, called \tool, that addresses the limitations above to triage Android access control policies to identify the attack operations  that adversaries are authorized to launch.  \tool addresses three key issues in multi-policy access control analysis.  First, the \tool design leverages the insight that access control policies whose decisions are combined via intersection can be analyzed independently~\cite{wcs+02}, enabling the addition of new policy models without impacting others.  Second, \tool accounts for how adversaries may
exploit discretionary elements in Android access control
to expand the permissions available to themselves and/or
victims, which we call {\em permission expansion}.  By accounting for permission expansion, \tool can detect attacks that are missed by analyses that use policies as configured.  Third, \tool reasons about how permissions may be exploited in attacks to convert unsafe data flows computed by past systems~\cite{jsz03} into the specific {\em attack operations} that can really be performed. 

To demonstrate the effectiveness of \tool to perform access control analysis on multiple Android access control policies, we investigate the impact of the 
recently introduced Scoped Storage defense~\cite{scopedstorage,lee21ieeesp}.
Android uses a separate filesystem partition for many dynamically processed applications files, including media and application updates, called the {\em external storage partition} for historical reasons, but this filesystem has been the source of many exploits, including Checkpoint vulnerability~\cite{MITD}.  We show how \tool enables the independent addition of the Scoped Storage policy model to the traditional Android access control model
to assess research questions about
the impact of this defense using eight freshly installed
Android releases: three Google Android versions and five
OEM Android versions. 
\tool shows that Scoped Storage reduces the number of possible attack operations in external storage across Google and OEM systems by over 50\%.  However, on OEM devices,
\tool shows that a small number of ``legacy apps'' that are not compliant with Scoped Storage are responsible for many attack operations, and even two previously unknown vulnerabilities.  We show how to use \tool to evaluate a what-if scenario where all apps are compliant with Scoped Storage, finding that this scenario reduces the number of attack operations by 12-28\% across versions, but reduces the number of adversaries that could launch such attack operations drastically (over 65\% for the OEM devices).  
\tool is
available as open source on Github\footnote{\tool Repository: \url{http://github.com/yxl74/PolyScope}}.  We have reported all vulnerabilities.

This paper makes the following contributions:


\begin{itemize}
\itemsep0.3em 
    \item \textit{We propose the \tool analysis tool to triage Android filesystem access control, accounting for permission expansion and specialized controls in Android systems.}
    \tool reduces Android's access control policies to the set of attack operations that adversaries are authorized to launch.
    
    \item \textit{We extend \tool to compute attack operations accounting for the newly added Scoped Storage defense.}
    We show how \tool can be extended to reason about Scoped Storage and other {\em restrictive} access control policies independently.
    
    \item \textit{We use \tool to triage eight Google and OEM Android releases to assess the effectiveness of Scoped Storage quantitatively.} We find that Scoped Storage reduces the number of attack operations significantly (54-71\%), but Scoped Storage is not fully employed, preventing more reduction (12-28\%) and exposing two vulnerabilities. 
\end{itemize}

The remainder of this paper is as follows.
Section~\ref{sec:motivation} motivates the need for more effective access control analysis.  
Section~\ref{sec:overview} overviews of the \tool's approach.
Section~\ref{sec:threat} defines our threat model.
Sections~\ref{sec:design} and~\ref{sec:impl} describe the design and implementation of \tool.
Section~\ref{sec:eval} evaluates how Scoped Storage impacts the filesystem threats across eight Android releases.
Section~\ref{sec:discussion} describes current limitations and how they may be addressed.
Section~\ref{sec:relwork} examines differences from related work.
Section~\ref{sec:conc} concludes.

 

\vspace{-0.1in}
\section{Motivation}

\label{sec:motivation}

In this section, we motivate the goals of our work by using an example to show the conditions when filesystem vulnerabilities may occur and then we outline our goal of assessing the impact of Scoped Storage to reduce attacks in external storage on Android devices.

\vspace{-0.1in}
\subsection{An Example Vulnerability}
\label{subsec:example}
\vspace{-0.05in}

A recent vulnerability discovered in Android services using the ContactsProvider allows untrusted apps to gain access to privileged files~\cite{cve202220220}.
The ContactsProvider enables services to retrieve files on behalf of apps by a URI specifying the location of a file.
An untrusted app may lure a service's ContactsProvider into using a maliciously crafted URI that resolves to a symbolic link created by the untrusted app.  Through this symbolic link, the untrusted app can access any file to which the service is authorized, which may include some privileged files.  This is an example of a {\em link traversal attack}. 

\begin{figure}[t]
\resizebox*{!}{3cm}{\includegraphics[width=2.3in]{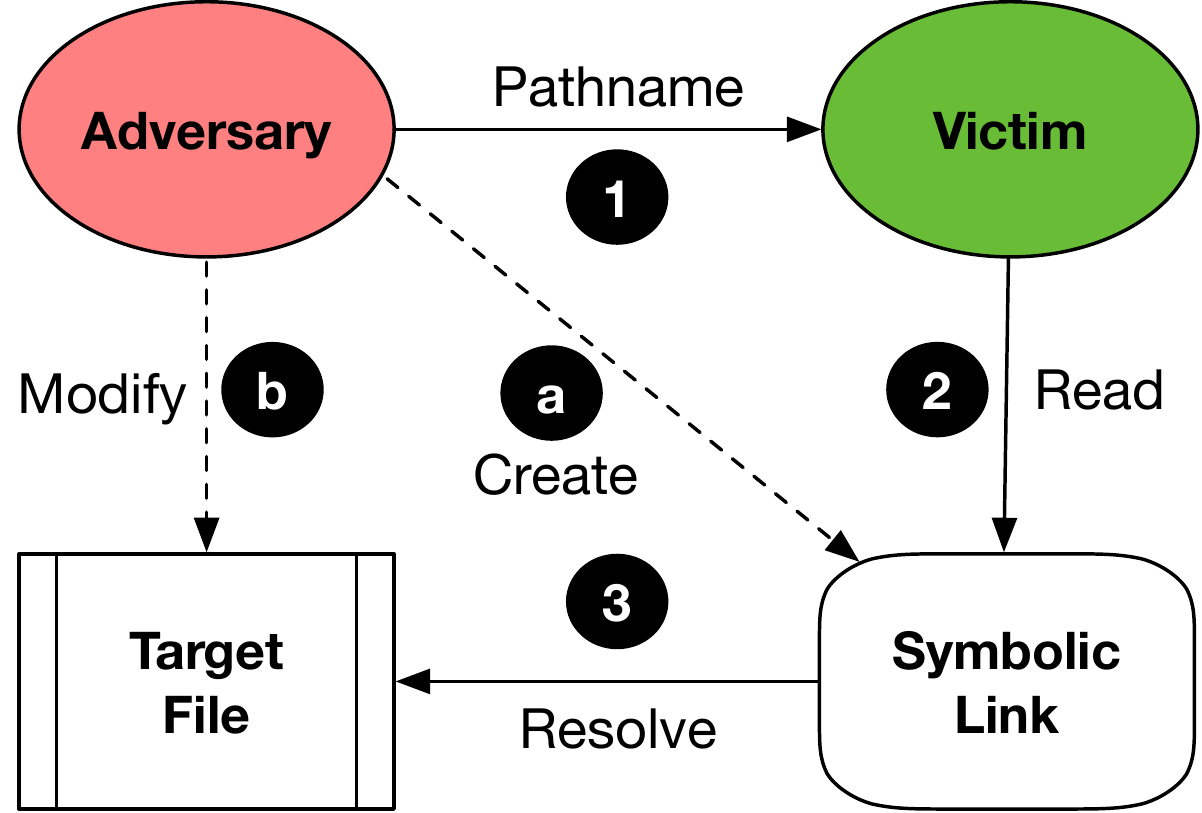}}
\centering
\caption{{\bf ContactsProvider Vulnerability}: (1) Adversary provides pathname to victim (as URI) to (2) lure the victim to an adversary-created symbolic link (a) that (3) the victim resolves to the target file enabling the adversary to modify the file indirectly through the victim (b).}
\label{fig:example}
\vspace{-0.1in}
\end{figure}


Figure~\ref{fig:example} shows exploitation of the vulnerability.  The adversary sends a request URI (Pathname in Figure~\ref{fig:example}) to the victim (service running ContactsProvider) {\footnotesize\circled{1}} that directs the victim to a symbolic link created by the adversary {\footnotesize\circled{a}}.  When the victim uses its read permission to the symbolic link  {\footnotesize\circled{2}}, the operating system resolves the link {\footnotesize\circled{3}} to return access to the target file.  This vulnerability may enable the adversary to leak, modify, and delete the target file {\footnotesize\circled{b}} to which the adversary normally lacks access.

This vulnerability occurred because adversaries of the service running ContactsProvider have the permission to create a symbolic link in a directory to which the service also has access.  The file squatting attack found by Checkpoint~\cite{MITD} that is described in the Introduction is caused by the same conditions, although in this case the adversary creates a file instead of a symbolic link in step {\footnotesize\circled{a}}. 
\vspace{-0.15in}
\subsection{The Android External Storage Problem}
\label{subsec:external}
\vspace{-0.05in}

Filesystem vulnerablities have been a particular problem in the external storage partitions\footnote{The external storage partition originally reflected a separate storage device (e.g., SD card), but modern Android systems now host the external storage partition on device storage.} of Android systems.  The Android external storage partition provides a filesystem for apps 
to store application-specific data.
As users may benefit from the ability of multiple apps to access the same data (e.g., to edit photos and other media generated by other apps), Android provides the ability for multiple apps to share access to external storage.  However, researchers have shown that vulnerabilities are caused when multiple mutually untrusting subjects can modify the same  directory~\cite{sting}, so many vulnerabilities have been found in external storage, such as the Checkpoint vulnerability~\cite{MITD}.

In Android 10, an experimental implementation of a new defense, called Scoped Storage, was introduced with the aim of providing improved protection to data on external storage~\cite{lee21ieeesp}. Scoped Storage classifies external storage into package-specific private and shared directories, each with limited access rights.  Files in private directories can only be accessed by apps associated with the package, whereas files in shared directories can only be accessed by other apps when they have the appropriate Android permissions.  Further, write permission to files in shared directories require special Android permissions (i.e., available only to approved packages) or user consent.  Details on how the Scoped Storage defense works are provided in Section~\ref{subsec:label}.


In this paper, the question that we examine is how to extend access control policy analysis to determine the impact that Scoped Storage has on preventing filesystem vulnerabilities.  
An additional challenge is created because not all apps will utilize the Scoped Storage defense.  Some apps may still operate under defenses prior to Scoped Storage, now called {\em legacy storage}.  An app that uses legacy storage has write privileges to others' files and permits other apps to write to the app's files, possibly introducing vulnerabilities.  Thus, we also assess the effect of using legacy storage on the security of the external storage.

\vspace{-0.15in}
\section{Background}
\label{sec:back} 
\vspace{-0.05in}

We describe access control policy analysis methods and describe limitations in current analysis techniques.

\vspace{-0.1in}
\subsection{Access Control Policy Analysis}
\label{subsec:ivs} 
\vspace{-0.05in}


Access control policy analysis~\cite{jaeger02sacmat,setools} computes the authorized information flows among subjects and objects from a system's access control policies.  An access control policy authorizes an {\em information flow from a subject to an object} if the policy allows that subject to perform an operation that modifies the object, called a {\em write-like operation}, and authorizes an {\em information flow from an object to a subject} if the policy allows that subject to perform an operation that uses the object's data, called a {\em read-like operation} (e.g., read or execute).  Some operations may be both read-like and write-like, enabling information flow in both directions.

However, modern Android systems have hundreds of thousands of access control rules, so there are many, many authorized information flows.  Thus, researchers developed access control analyses to identify secrecy problems~\cite{chen17acsac,enck09ccs,wae+17,Wang2015,aafer15ccs} and integrity problems~\cite{jaeger03usenix,chen09ndss}.  The vulnerability in Section~\ref{subsec:example} is an example of an integrity problem, where an adversary controls a filesystem resource used by a victim to perform the attack.  
To detect integrity problems, access control analyses are inspired by integrity models, such as Biba integrity~\cite{b77}, to detect information flows from adversary processes to victim processes.  Such information flows are called {\em integrity violations} (IVs), which are defined more formally as a tuple of resource, adversary, and victim, where the access control policy authorizes an information flow from the adversary to the resource (i.e., the adversary is authorized to perform a write-like operation on the resource) and authorizes an information flow from the resource to the victim (i.e., the victim is authorized to perform a read-like operation on the resource). 

\begin{figure*}[t]
\includegraphics[width=6.5in]{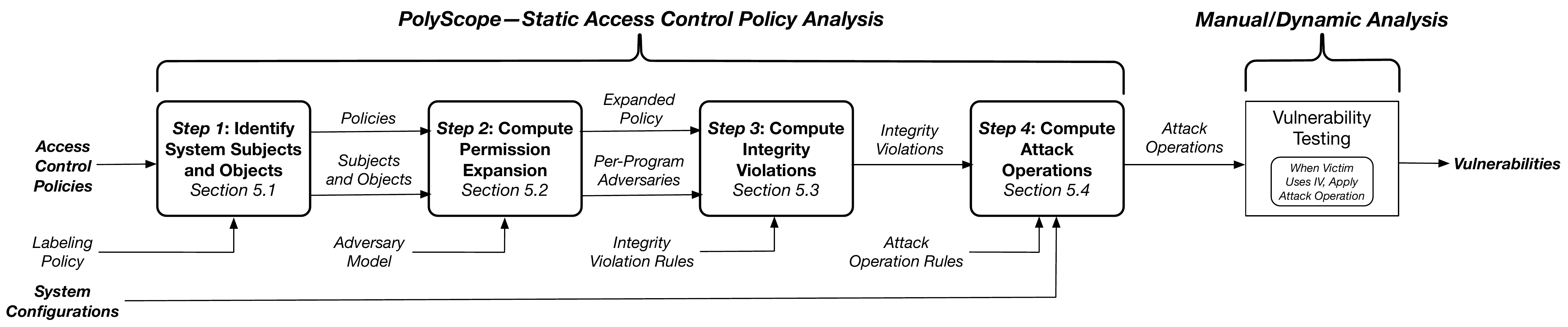}
\centering
\caption{{\bf \tool Logical Flow:} \tool computes per-subject adversaries (Step 1), permission expansion by those adversaries (Step 2), the integrity violations to which adversaries are authorized (Step 3), and the attack operations adversaries may perform to launch attacks (Step 4) as test cases for vulnerability testing.}
\label{fig:flow}
\vspace{-0.1in}
\end{figure*}

\vspace{-0.1in}
\subsection{Limitations of Current Techniques}
\label{subsec:limits}
\vspace{-0.05in}

Access control policy analyses attempt to solve three main problems to help identify vulnerabilities, but current approaches suffer from key limitations on each problem.

The first problem is to {\bf characterize subjects and objects properly given multiple access control policies}. Traditionally, research only considered one access control policy at a time.  Researchers have recently proposed techniques to reason about MAC and DAC policies in combination~\cite{chen17acsac,BigMAC}, but they have not considered how to compose subjects and objects from multiple policies systematically. As Android includes new access control mechanisms, such as Scoped Storage, how to characterize subjects and objects accurately across multiple policies becomes challenging. 
The second problem in using access control policy is to {\bf identify the permissions that adversaries could control to launch attacks comprehensively}. 
A problem is that adversaries may exploit the flexibility in policy models like DAC and Android permissions to add permissions that create more integrity violations. For example, adversaries may obtain Android permissions from unsuspecting users or adversaries may grant permissions to object they "own" to potential victims to lure them into attacks.  Researchers have previously identified problems caused by DAC policy flexibility~\cite{hru76,flask-inevitability} that limit its ability to prevent unauthorized access. While in theory MAC policies could be configured to block changes from creating integrity violations, MAC policies are more complex to configure and are unforgiving if a needed permission is not granted.  

The third problem is to {\bf compute the operations that an adversary may be authorized to employ to launch attacks}, which we call {\em attack operations}.  Once we know that an adversary has been authorized permissions that create an integrity violation, a question is how an adversary may exploit those permissions to launch attacks.  While integrity violations are a necessary precondition for attacks, adversaries must be able to perform the operations necessary to launch attacks.  Android systems prevent attack operations in some cases, such as prohibiting the use of symbolic links in external storage.  

\vspace{-0.1in}
\section{\tool Overview}
\vspace{-0.05in}
\label{sec:overview}


In this paper, we present a new Android access control analysis tool, called \tool, that computes the set of authorized attack operations for an Android system while overcoming the limitations described above.  \tool takes the {\em access control policies} and {\em system configurations} for the particular system under test and produces a set of {\em attack operations} that should be vetted in vulnerability testing. 
Figure~\ref{fig:flow} shows \tool's approach, where the two user (analyst) inputs, Android access control policies and Android system configurations, are highlighted in bold\footnote{The specific inputs are described in "Data Collection" in Section~\ref{sec:impl}.}.  The sources of other inputs are described below.  

In Step 1, 
\tool maps processes and filesystem resources to unique subjects and objects for the access control policies using a  {\em labeling policy}.  
Since Android access control combines policies in a restrictive manner (i.e., all policies must authorize an operation), labeling can be done independently for each policy model, which makes it straightforward to extend Android access control with Scoped Storage as described in Section~\ref{subsec:label}.  
In Step 2, \tool determines the permissions that may be associated with subject and object labels by modeling how each subject's adversaries may expand the permissions available to themselves and their victims by exploiting the flexibility in Android and DAC access control policies, as described in Section~\ref{subsec:expand}.  Adversary models are also chosen once for the combination of policies, where we show the adversary model we use for Android systems in the Threat Model in Section~\ref{sec:threat}.  In Step 3, \tool uses these expanded permissions to compute integrity violations based on {\em integrity violation rules} defined in Section~\ref{subsec:ivdef}.  In Step 4, \tool uses these integrity violations to compute the attack operations possible using {\em attack operation rules} that reference additional {\em system configurations} as defined in Section~\ref{subsec:operations}.  These rule sets are defined by \tool and are independent of the policy model.


\tool computes the attack operations to triage Android releases for  vulnerabilities authorized by access control policies.  
Attack operations computed in Step 4 identify all the operations that adversaries are capable of performing to modify resources to launch attacks.  
Using the computed attack operations, an analyst can perform vulnerability testing on victim applications either manually or preferably using dynamic analysis.  
In Section~\ref{sec:impl}, we describe a basic dynamic analysis analysis method to detect victim use of resources that can be modified by attack operations, from which we find two new vulnerabilities from subsequent manual testing in Section~\ref{subsec:case}.  

\vspace{-0.1in}
\section{Threat Model}
\label{sec:threat}
\vspace{-0.05in}
In this paper, we assume that adversaries may modify any part of the filesystem 
to which they are authorized by the combination of Android access control policies, which \tool computes as {\em integrity violations}.  In addition, we assume that adversaries will perform any operation to launch attacks on integrity violations that are possible given the system configuration, called {\em attack operations}.  In this section, we examine the specific types of integrity violations and attack operations, we consider as threats in this paper.

\begin{figure}[t]

\resizebox*{!}{3cm}{\includegraphics[width=2.3in]{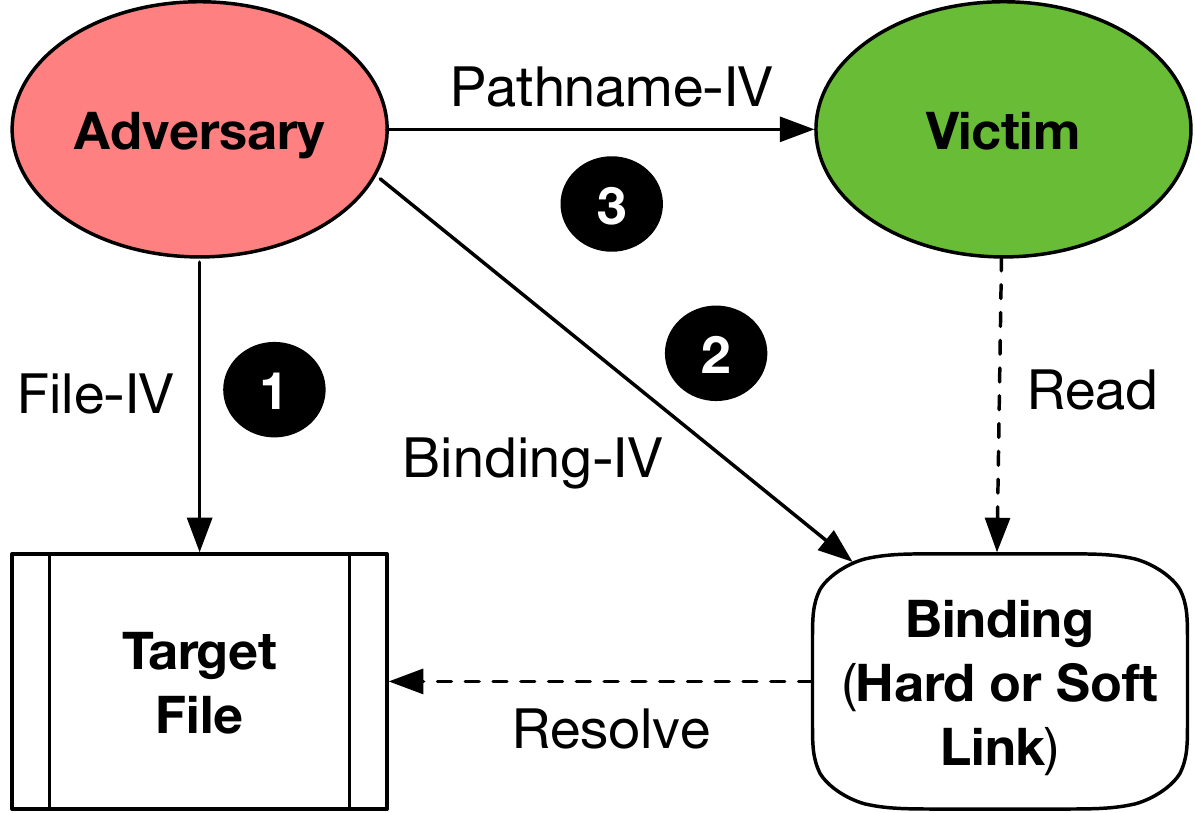}}
\centering
\caption{{\bf Integrity Violation (IV) Classes}: (1) {\em File-IVs} grant adversaries direct access to modify files that victims use; (2) {\em Binding-IVs} grant adversaries the ability to modify name resolution of file names; and (3) {\em Pathname-IVs} enable adversaries to lure victims to the part of the filesystem they can modify.}
\label{fig:classes}
\vspace{-0.1in}
\end{figure}


Based on the example in Figure~\ref{fig:example}, we show the three classes of integrity violations (IVs) we consider in developing \tool in Figure~\ref{fig:classes}.
on filesystem access, covering a wide variety of vulnerabilities including confused deputy~\cite{confused-deputy} and time-of-check-to-time-of-use (TOCTTOU) vulnerabilities~\cite{mcphee74,bishop-dilger}.  Related to Figure~\ref{fig:example}, we show these integrity violation classes in Figure~\ref{fig:classes}.  First, {\em file-IVs} allow adversaries to modify target files that are authorized to victims directly {\footnotesize\circled{1}}, possibly leading victims to unexpected use of adversary-controlled data.  File-IVs may be distinguished further by whether the victim can read ({\em read-IVs}), write ({\em write-IVs}), and/or execute ({\em exec-IVs}) the IV file.  Second, {\em binding-IVs} enable adversaries to redirect victims to target files during name resolution {\footnotesize\circled{2}}, causing victims to operate on files chosen by adversaries.  Third, {\em pathname-IVs} enable adversaries to lure victims to an adversary-controlled part of the filesystem using an adversary-supplied pathname {\footnotesize\circled{3}}, which is the integrity violation exploited in the example vulnerability of Section~\ref{subsec:example}. 


For each integrity violation found, we assume that an adversary may attempt any possible attack operation.  
File-IV attack operations simply {\em modify the resource} awaiting use (read, write, or execute) by the victim.  Binding-IV attack operations direct the victim to a resource chosen by the adversary, using link traversal or file squatting attacks.  A {\em link traversal} attack directs a victim to access a resource to which the adversary is not authorized.  A {\em file squatting} attack plants an adversary-controlled resource that a victim may use. 
Pathname-IV attack operations lure a victim who processes adversary-controlled pathnames (e.g., URLs via IPCs) to an adversary-controlled binding to exploit a link traversal, which we call a {\em luring traversal}.


In developing \tool, we assume trust in some components of Android systems.  First, we assume that the Android operating system operates correctly, including enforcement of its access control policies and system configurations.  For example, we trust the Android operating system to satisfy the reference monitor concept~\cite{jaeger_refmon}.  
Second, our assumptions about trust among user-space processes is determined by Google's Process Privilege Levels shown in Table~\ref{table:levels}.  A subject trusts services/apps at its privilege level or higher.  Other subjects are adversarial.  We have shown how to validate that the Android policy is consistent with the trust implied by these privilege levels~\cite{lee21usenix}.
\begin{table*}[ht]
\centering
\resizebox*{!}{3.5cm}{
\begin{threeparttable}%
    \caption{Google's Process Privilege Levels~\cite{PPRIV_LEVEL}}
    \small
\begin{tabular}{c|l}
 \textbf{Process Level$^1$} &\textbf{Level Membership Requirements}\\  \midrule
Root Process (T5)    & Process running with UID root (e.g., MAC labels {\tt kernel} and {\tt init})       \\
System Process (T4)   & Process running with UID system (e.g., MAC label {\tt system server})       \\
Service Process  (T3)   & AOSP core service providers (e.g., MAC labels {\tt bluetooth} and {\tt mediaserver})      \\
Trusted Application Process (T2)    & AOSP default and vendor apps (e.g., MAC labels {\tt platform\_app} and {\tt priv\_app})      \\
Untrusted Application Process (T1)     & Third-party applications (e.g., MAC label {\tt untrusted\_app})        \\
Isolated Process  (T0) &  Processes that are expected to receive adversarial inputs (e.g., MAC label {\tt webview})         \\
\end{tabular}
\begin{tablenotes}\scriptsize
    \item[1] Listing types of processes based on their privilege level, from high to low with root processes being most privileged (T5) and isolated processes being the least privileged (T0).  We group T0 and T1 together calling the resultant level T1 in the evaluation in Section~\ref{sec:eval}.
\end{tablenotes}
\label{table:levels}
\end{threeparttable}%
}
\end{table*}

\vspace{-0.13in}
\section{\tool Design}
\label{sec:design}
\vspace{-0.05in}


In this section, we examine the design challenges in computing attack operations for Android systems.  In particular, 
we focus on four key steps outlined in the \tool overview in Section~\ref{sec:overview}.  

\vspace{-0.1in}
\subsection{Identify Subjects and Objects}
\label{subsec:label}
\vspace{-0.05in}


The first step is to identify the subjects and objects in the system based on the combination of access control policies.   Access control determines when a process may perform an operation on a system resource.   However, access control policies are typically expressed in terms of identifiers for subjects and objects (e.g., user IDs and group IDs), rather than individual processes and resources, to enable access control decisions to be made as the system's processes and resources evolve dynamically.   The mapping of access control identifiers to system processes and resources forms a {\em labeling policy}.
Since Android uses a combination of access control policies, a question is how to determine the combination of identifiers (i.e., the labeling policy( that constitute subjects and objects for this combination of policies.  


Researchers have previously found it useful to identify combinations of policies that are either restrictive or authoritative~\cite{wcs+02}.  A composition of policies is said to be {\em restrictive} when any policy can deny access (i.e., the authorized permissions are the {\em intersection} of each policy's permissions).  On the other hand, a combination of  policies is said to be {\em authoritative} if any policy may grant a permission, even if it is denied by another policy (i.e., the authorized permissions are the {\em union} of each's authorized permissions).
In either case, the subjects and objects for each policy can be determined independently and the resultant permissions are a simple composition (e.g., intersection or union) of the permissions assigned to the individual policies (i.e., between each policies' subjects and objects).

We find that the combination of Android access control policies is restrictive.  Android requires that all policies must authorize an operation for it to be permitted. As a result, we can simply extend the subjects and objects in \tool by specifying the Scoped Storage subjects and objects identifiers independently from the other Android policies.  

The subject and object identifiers used by \tool for the Android access control policies, other than Scoped Storage, are as follows\footnote{For details justifying these choices, see the original \tool paper~\cite{lee21usenix}.}.

\begin{itemize}
\item {\bf SELinux Type Enforcement (TE)}: Subjects and objects are assigned TE labels.
\item {\bf SELinux Multilevel Security (MLS)}: Subjects and objects are assigned MLS category sets.
\item {\bf UNIX Discretionary Access Control (DAC)}: Subjects are assigned DAC User ID and a set of DAC groups (i.e., an owner Group ID and supplemental group IDs).
Objects are assigned an DAC UID and DAC GID for its owner. 
\end{itemize}

The Scoped Storage defense (see Section~\ref{subsec:external}) has been designed to control access to each app's files in the Android external storage partition.  Recall that Scoped Storage separates app-specific storage into private and shared directories.  Private directories may only ever be accessed by the owning app, but other apps may gain access to shared directories through Android permissions or user consent.  The apps may obtain read (only) access to any shared directories thought the {\tt READ\_EXTERNAL\_STORAGE} (REX) Android permission.  Write access may be obtained in two ways.  First, an app may gain write access to files in a shared directory by obtaining user consent through the Storage Access Framework (SAF). Second, apps that are vetted prior to publication in the Google Play Store are eligible to obtain the {\tt MANAGE\_EXTERNAL\_STORAGE} (MES) permission that grants them full read and write access to external storage.

For compatibility purposes, some apps may be declared as {\em legacy} apps, which basically means that these apps use the access controls that pre-date Scoped Storage.  Legacy apps may gain read and write access to any package's shared directories using the {\tt WRITE\_EXTERNAL\_STORAGE} (WEX, deprecated since Android 11) permission. In addition, legacy apps are allowed to place files in the root directory of external storage, which is shared among legacy apps and other system services. This root directory is not accessible to apps compliant with Scoped Storage. 


As a result, the Scoped Storage access control policy governs access based on the app's package name and ad hoc choices of Android permissions.  Specifically, Scoped Storage subjects and objects are defined below.  

\begin{itemize}
    \item {\bf Scoped Storage Subject:} Each subject is defined as a combination of: (1) the package name associated with an app; (2) whether the app opts for Scoped Storage (i.e., is legacy or not); and (3) the app's Scoped Storage Android permissions (i.e., REX and MES). 
    \item {\bf Scoped Storage Object:} Each object is defined by: (1) the package name associated with the Scoped Storage directory; (2) private or shared directory; and (3) any ad-hoc user consent. 
\end{itemize}

Because access control policies in Android are restrictive, we can add the subject and object definitions for Scoped Storage independently from other policies.  
In analysis, \tool only needs to intersect the outcome of the authorization using the Scoped Storage policy with that of the MAC (TE and MLS) and DAC authorization results to determine which subjects can access an object.

\vspace{-0.15in}
\subsection{Compute Permission Expansion}
\label{subsec:expand}
\vspace{-0.05in}

A key difficulty for OEMs is predicting which resources may be accessible to adversaries and victims to derive attack operations accurately.  A problem is that while MAC policies are essentially fixed (i.e., between software updates), DAC permissions may be modified by adversaries to increase the attack operations that they could execute.   We identify two ways that adversaries may modify permission assignments on Android systems: (1)~adversaries may obtain Android permissions that augment their own DAC permissions and (2)~adversaries may delegate DAC permissions for objects that they own to potential victims.  First, some Android permissions granted by users allow adversaries to gain new DAC permissions to access additional resources that may enable attacks.  Second, by delegating DAC permissions to objects they own, adversaries may also lure potential victims to resources that adversaries control.  

\paragraph{Adversary Permission Expansion}
In Android systems, some Android permissions are implemented using DAC groups.  As described above, a process is associated with a single UID and GID, but also an arbitrarily large set of supplementary groups that enable further "group" permissions.  Thus, when a user grants an Android permission associated with one or more DAC groups to an app, there is a direct expansion of that app's permissions in terms of its DAC permissions.  Since the MAC policies are generally lenient in Android systems, these new DAC permissions may grant privileges that enable attacks.  For \tool, we assume that subjects can obtain all of their "normal" Android permissions and are granted all of their "dangerous" permissions by users for analysis, as described in the previous section. Scoped Storage adds another kind of adversary permission expansion by allowing apps to declare themselves as legacy apps. The legacy flag grants write privilege to files in multiple locations of external storage and greatly boosts an adversary's capability to launch attacks. 
One of the vulnerability case studies we highlight in Section~\ref{subsec:case} exploits the use of the legacy flag for adversary permission expansion.

\paragraph{Victim Permission Expansion}
Researchers have long known that allowing adversaries to administer DAC permissions for their own objects can present difficulties in predicting possible permission assignments.  Researchers proved that the {\em safety problem} of predicting whether a particular unsafe permission will ever be granted to a particular subject for a typical DAC protection system, i.e., an access matrix for subjects and objects where objects and permissions may be added in a single command, is undecidable in the general case~\cite{hru76}.  
As a result, researchers explored alternative DAC models for which the safety problem could be solved, such as the take-grant model~\cite{take-grant}, the typed access matrix~\cite{tam}, and policy constraints~\cite{tidswell00ccs}.

Using the ability to manage DAC permissions to objects they own, adversaries can grant permissions to their resources to victims, expanding the set of resources that victims may be lured to use.  In many cases, victims have MAC permissions that grant them access to adversary directories, but vendors use DAC permissions to block access.  However, when adversaries own these directories, they can simply grant the removed permissions to potential victims.  

\paragraph{Scoped Storage Impact on Permission Expansion}
Scoped Storage permits two kinds of adversary permission expansion.  First, apps that can obtain the MES permission can modify any file in external storage.  Fortunately, Google must vet any app before it can even request that permission, but even some vetted apps only have a T1 Google privilege level (see Table~\ref{table:levels}), exposing some risks.  
Legacy apps 
obtain write permissions to shared directories by default, which is likely a bigger risk. 
Second, apps can request write access to files from users.  While users may grant access to any file in a shared directory, in general, the impact of Scoped Storage is largely bypassed if that is done comprehensively.  In this work, we do not apply user grants of individual files in shared directories to permission expansion.  Studying possible risks of such user grants is future work. 

For pre-Scoped Storage systems, we assume that victims can expand permissions (i.e., perform adversary expansion) to obtain the REX/WEX permissions since most apps need access to shared folders in external storage. However, for post-Scoped storage systems, adversaries cannot cause any form of victim permission expansion because they cannot change the permissions for other users, and victim apps no longer need to declare REX/WEX to access shared locations in external storage. Outside of external storage, both types of permission expansion threats still remain.

\vspace{-0.12in}
\subsection{Compute Integrity Violations}
\label{subsec:ivdef}
\vspace{-0.05in}

We show how to compute integrity violations for file-IVs, binding-IVs, and pathname-IVs defined in Section~\ref{sec:threat}.  


\paragraph{Computing File Integrity Violations}
A file integrity violation occurs when a victim subject may use (i.e., has permission to read, write, or execute) a file object to which an adversary subject has permission to modify.  In practice, many subjects read file objects their adversaries may write ({\tt read-IVs}), but even risks are greater if the subject executes ({\tt exec-IVs}) or also modifies such files ({\tt write-IVs}).  For exec-IVs, executing input from an adversary enables an adversary to control a victim's executable code.  For write-IVs, if a subject writes to a file object its adversaries also may write, then adversaries may be able to undo or replace valid content.

\vspace{0.05in}
\begin{scriptsize}
\begin{minted}[]{C}
{read|write|exec}(file, victim) &&  // victim can access file,
adv-expand(file, adversary) &&      // but adv-expanded perms 
write(file, adversary)              // enables to modify file
    -->
{read|write|exec}-IV(file, victim, adversary)
\end{minted}
\end{scriptsize}

This rule determines whether the victim is authorized by the combination of access control policies for reading, writing, or executing file objects, using the {\tt \{read|write|exec\}} predicate.  The rule accounts for the adversary's expansion of their own permissions, as indicated by the predicate {\tt adv-expand}.  If the adversary also has write permission to the file object ({\tt write} predicate), then the associated integrity violation is created. 

\paragraph{Computing Binding Integrity Violations}
A binding integrity violation occurs when a subject may use a binding object that adversaries can modify in resolving a file pathname.  

\vspace{0.05in}
\begin{scriptsize}
\begin{minted}[]{C}
use(binding, victim) &&          // victim can use binding,
adv-expand(file, adversary) &&   // but adv-expanded perms
write(binding, adversary)        // enable to modify binding
    -->
binding-IV(binding, victim, adversary)
\end{minted}
\end{scriptsize}

This rule parallels the rule for file-IVs, except that this rule applies to a victim having the permission to use a binding object in name resolution ({\tt use} predicate). 

\paragraph{Computing Pathname Integrity Violations}
Pathname integrity violations are binding integrity violations that are possible when a subject uses input from an adversary to build a file pathnames used in name resolution.  First, adversaries must be authorized to communicate with the victim.  Second, through their input, adversaries can lure victims to any bindings they choose, enabling them to expand the IVs available for exploitation by victim permission expansion.

\vspace{0.05in}
\begin{scriptsize}
\begin{minted}[]{C}
write(ipc, adv, vic) &&           // may send IPCs to victim
vic-expand(binding, adv, vic) &&  // and expand victim perms
binding-IV(binding, vic, adv)     // to lure victim
    -->
pathname-IV(binding, vic, adv)
\end{minted}
\end{scriptsize}

Adversaries must be granted write privilege to communicate to the victim, as defined in the {\tt write} predicate.  Android services may use Binder IPCs, and we further limit {\tt write} to use IPCs that communicate URLs for Android services.  The adversary can use victim expansion to increase the set of bindings the victim is authorized to use by {\tt vic-expand}.  If that binding object meets the requirements of a binding-IV (see above), then a pathname-IV is also possible for this victim. 

\vspace{-0.12in}
\subsection{Compute Attack Operations}
\label{subsec:operations}
\vspace{-0.03in}

We define how \tool computes attack operations from the integrity violations produced in the last section and the relevant system configurations.  We identify four types of attack operations that an adversary could use to exploit the three types of integrity violations: (1) file modification for file IVs; (2) file squatting for binding-IVs; (3) link traversal for binding-IVs; and (4) luring traversal for pathname-IVs.  

\paragraph{File Modification Attacks}
For read/write/exec IVs, the attack operation is to modify the objects involved in each IV.  However, Android uses some read-only filesystems, so not all files to which adversaries have write privilege are really modifiable.  Thus, the rule for {\em file modification} operations additionally checks whether the file is in a writable filesystem.

\vspace{0.05in}
\begin{scriptsize}
\begin{minted}[]{C}
{read|write|exec}-IV(file, victim, adversary) &&
fs-writable(file)      // file's filesystem is writable
    -->
file-mod(file, victim, adversary)
\end{minted}
\end{scriptsize}

\paragraph{File Squatting Attack}
In a file squatting attack, adversaries plant files that they expect that the victim will access.  The adversary grants access to the victim to allow the victim to use the adversary-controlled file.  This attack operation gives the adversary control of the content of a file that the victim will use.  To perform a file squatting attack operation, the adversary must really be able to write to the directory to plant the file.  Thus, the rule for {\em file squatting} operations is essentially the same as for file modification, but applies to binding-IVs.  

\vspace{0.05in}
\begin{scriptsize}
\begin{minted}[]{C}
binding-IV(binding, victim, adversary) &&
fs-writable(binding)     // binding's filesystem is writable
    -->
file-squat(binding, victim, adversary)
\end{minted}
\end{scriptsize}

In this rule, we assume that the adversary predicts the filenames used by the victim.  In the future, we will explore extending the rule to account for that capability.

\paragraph{Link Traversal}
A link traversal attack is implemented by planting a symbolic link at a binding modifiable by the adversary, as described in Section~\ref{subsec:example}.  However, Android also uses some filesystem configurations that prohibit symbolic links, so not all bindings to which adversaries have write privilege and are in writable filesystems allow the creation of the symbolic links necessary to perform link traversals.  Thus, the rule for {\em link traversal} operations extends the rule for file squatting to account for this additional requirement.  

\vspace{0.05in}
\begin{scriptsize}
\begin{minted}[]{C}
binding-IV(binding, victim, adversary) &&
fs-writable(binding)     // binding's filesystem is writable
symlink(binding) &&      // and allows symlinks
    -->
link-traversal(binding, victim, adversary)
\end{minted}
\end{scriptsize}


\begin{figure*}[t]
\includegraphics[width=6.0in]{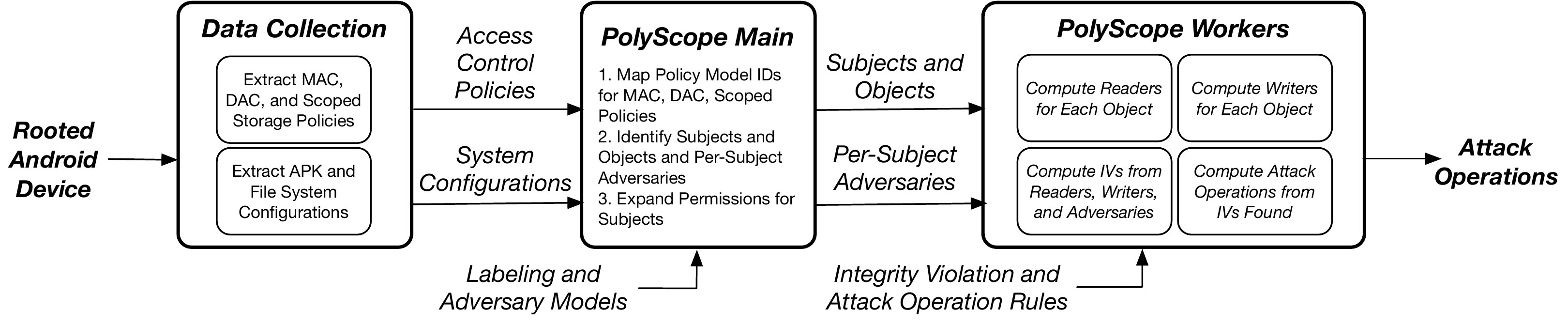}
\centering
\caption{{\bf \tool Implementation:} \tool collects inputs from rooted Android devices to identify subjects and objects.  IVs and attack operations are computed in parallel per object by \tool workers.}
\label{fig:impl}
\vspace{-0.1in}
\end{figure*}

\paragraph{Luring Traversal}
An adversary may lure a victim to a binding controlled by the adversary to launch an attack operation.  However, the Android FileProvider class can prevent such attacks.  Specifically, the FileProvider class requires that clients open files themselves and provide the FileProvider with the resultant file descriptor.  Since the clients open the file, they perform any name resolution, so the potential victim is no longer prone to pathname vulnerabilities.  Thus, the rule for {\em luring traversal} operations extends the rule for link traversal for pathname-IVs by requiring the absence of any FileProvider class usage.  OEMs still have many services and privileged apps that do not employ the FileProvider class, leaving opportunities for pathname-IVs to be attacked.

\vspace{0.05in}
\begin{scriptsize}
\begin{minted}[]{C}
pathname-IV(binding, victim, adversary) &&
fs-writable(binding) &&    // binding's filesystem is writable
symlink(binding) &&        // and allows symlinks
no-file-provider(victim)   // victim does not use FileProvider
    -->
luring-traversal(binding, file, victim, adversary)
\end{minted}
\end{scriptsize}


While it is possible that the victim has implemented an extra defense in Android middleware (e.g., Customized Android Permission) to prevent IPCs, we do not yet account for these defenses.  Including these defenses is future work.


\vspace{-0.12in}
\section{Implementation}
\label{sec:impl}


The \tool tool is implemented fully in Python in about 3300 SLOC and is compatible with Android version 5.0 and above.  
The \tool implementation is shown in Figure~\ref{fig:impl}. First, \tool extracts access control policies and system configurations automatically in a Data Collection phase.  Then, \tool's Main phase identifies all the subjects and objects as described in Section~\ref{subsec:label}, determines the per-subject adversaries according to the Google Process Privilege Levels in Table~\ref{table:levels}, and expands subject permissions as described in Section~\ref{subsec:expand}. Next, \tool Workers computes IVs as described in Sections~\ref{subsec:ivs} and~\ref{subsec:operations}, respectively.  We are able to parallelize these steps per object which has a significant performance impact, as described in Section~\ref{subsec:perf}. 
Additional implementation details are the same as for the original \tool tool~\cite{lee21usenix}.

\paragraph{Data Collection} \tool has a variety of data collection scripts to collect access control polices (i.e., MAC, DAC, Android permissions, and Scoped Storage) and system configurations (i.e., filesystem settings and FileProvider use) to provide inputs to \tool.  The methods are relatively straightforward for accessible files and processes, as described previously~\cite{lee21usenix}.  However, \tool scripts are not authorized to access all files, particularly those owned by root, so we run these scripts on rooted phones.  Recent work by Hernandez et al.~\cite{BigMAC} is able to extract MAC policy and part of DAC configuration from Android firmware images without rooting devices. However, this approach cannot extract all files located in some directories like {\tt /data}. As shown in Table 1 of their paper~\cite{BigMAC}, about 75\% of the files' DAC configuration in {\tt /data} cannot be retrieved, which we extract with our scripts.

Data collection for Scoped Storage requires collecting access control information for each package and Scoped Storage directory.  To collect packages and their Android permissions,
\tool queries the PackageManager service for all the APKs on the device. Then, \tool parses the Android manifest files for the extracted APKs to obtain the permission mapping. For Scoped Storage directories, \tool extracts the Scoped Storage database file owned by MediaProvider to retrieve the owner of each external storage resource by package name.  
\tool collects the relevant program configurations (i.e., whether the victim includes a recommended defense, the FileProvider class) by reverse engineering the application's APK package to detect the presence of the FileProvider class.

\paragraph{\tool Main} This \tool component controls the steps in the \tool analysis.  As shown in Figure~\ref{fig:impl}, the \tool Main component runs three computations in series to identify subjects and objects and expand permissions, as described in Sections~\ref{subsec:label} and~\ref{subsec:expand}, respectively.  The most difficult step is to map the subjects between MAC (labels), DAC (UIDs/GIDs), and Scoped Storage (packages) policies. To find the mapping between UIDs and packages, \tool parses the {\tt package.list} file. However, we found that some package name-to-UID mappings are not one-to-one, as we expected and saw for MAC-to-DAC mappings, as multiple package names can be mapped to the same UID.  In this case, \tool over-approximates the mapping by assigning the union of all the package names that map to this UID for the subject.

\paragraph{\tool Workers}
Using the set of subjects, their adversaries, and objects, \tool can now compute the attack operations. The \tool implementation decomposes this computation into discrete components that can be parallelized, as shown in Figure~\ref{fig:impl}.  First, workers compute the subjects that can read (i.e., read and execute) and write each object.  This computation run per policy model (DAC, MAC, Scoped Storage) and the results per object are intersected.  Given the readers and writers for each object and policy model, the IVs for each object can be computed in parallel, one object per worker to roughly balance the load per worker.  Finally, the attack operations that applyfor each object's IVs can be computed in parallel as well. 
We validated the attack operations found can be performed and found no discrepancies.


\begin{table*}[t]
\centering
\resizebox{\textwidth}{!}{
\begin{threeparttable}[b]
    \caption{Summary of Integrity Violations (IVs) and Attack Operations Total and in External Storage across Vendor Releases}
    \label{tab:sum_table}
     \footnotesize
\begin{tabular}{@{}c|ccc|ccccc@{}}
                                 & \multicolumn{3}{c|}{\small{\textbf{Google Devices}}} & \multicolumn{5}{c}{\small{\textbf{OEM Devices}}} \\
                                 & \textbf{{\em Pixel3a 9.0}} &\textbf{Pixel3a 11.0} & \textbf{Pixel3a 12.0}  & \textbf{{\em Galaxy S20 9.0}} & \textbf{Galaxy S20 11.0} & \textbf{Galaxy S20 12.0}  & \textbf{Oneplus8T 11.0} & \textbf{Oneplus8T 12.0} \\ \midrule
\textbf{Total IVs}     & 2,124 & 1,334                   & 1,480                & 31,489             &12,713        &6,808  &  11,987 & 14,704       \\
\textbf{Total Attack Ops} & 2,512 & 1,628 & 1,794 & 36,258 & 15,414 & 8,465 & 11,540 & 14,135 \\ \hline
\textbf{Ext IVs}     &1,021 (48\%)                   & 260 (19\%)               & 374 (25\%)         &12,679 (40\%)      & 3,713 (29\%)                & 2,288 (34\%) & 4,365 (36\%) & 5,532 (38\%)   \\
\textbf{Ext Attack Ops}     & 527 (21\%)                  & 241 (14\%)              & 219 (12\%)           &11,336 (31\%)        &3,219 (21\%)          & 1,906 (23\%) & 3,929 (34\%) & 4,454 (32\%)  \\
\end{tabular}%
\begin{tablenotes}\scriptsize
    \item[] {\bf IVs} $= \sum_v^V |IV_{obj}(v)|$, where $IV_{obj}(v)$ returns the set of objects in the IVs for a victim $v \in V$ as computed per Section~\ref{subsec:ivdef}.
    \item[] {\bf Attack Ops} $= \sum_{iv}^{{\bf IVs}} |OP(iv)|$, where $OP(iv)$ returns the set of attack operations for an integrity violation $iv \in$ {\bf IVs}  computed per Section~\ref{subsec:operations}.
    \item[] Ext IVs (Attack Ops) are IVs (Attack Ops) whose objects are located in an external storage partition.
\end{tablenotes}
\end{threeparttable}%
}
\end{table*}

\paragraph{Testing for Vulnerabilities}  The ultimate goal is to determine whether the victim is vulnerable to any  of the attack operations.  However, a key challenge is to determine whether and when a victim may actually access a resource associated with an attack operation.  Just because a potential victim may be authorized to use a resource, does not mean it ever uses that resource. 
The major challenge is to drive the victim subjects' programs to cause all file usage operations, akin to fuzz testing.  Developing a fuzz testing approach for file operations is outside the scope of this paper, so we simply drive programs with available tools: (1) Android Exerciser Monkey; (2) Compatibility Testing Suite (CTS); and (3) Chizpurfle ~\cite{chizpurfle}. We use the Android Exerciser Monkey and CTS to emulate normal phone usage, and Chizpurfle to drive Android system services.  With this approach, we are able to find the vulnerabilities described in Section~\ref{subsec:case} manually. We discuss the challenges in automating vulnerability testing in Section~\ref{sec:discussion}.





\vspace{-0.1in}
\section{Evaluation}
\label{sec:eval}
\vspace{-0.05in}
In this section, we focus on measuring the impact of the Scoped Storage defense on threats to Android systems.  Note that we evaluated \tool against prior systems~\cite{BigMAC} in our previous paper~\cite{lee21usenix}.  Here, we apply the updated \tool tool described in this paper to six fresh installs of Android releases  that employ Scoped Storage\footnote{Oneplus8T is a relatively new phone that does not have any pre-Scoped Storage firmware available} (version 11.0 and above) and two fresh installs of Android releases that do not employ Scoped Storage (version 9.0). 

We explore the following research questions:

\begin{itemize}
    \item {\bf RQ1}: What fraction of the total number of threats in Android systems occur in external storage before and after the addition of Scoped Storage?
    \item {\bf RQ2}: How does Scoped Storage impact the types of integrity violations and attack operations that may be attempted in external storage?
    \item {\bf RQ3}: How many of the threats in external storage are due to legacy apps?
    \item {\bf RQ4}: How many attack operations and attackers could be removed if all legacy
apps are converted to Scoped Storage apps?
\end{itemize}

We first examine the distributions of IVs and attack operations within systems at-large and for external storage only (RQ1-RQ2).  These analyses show that Scoped Storage has reduced the number of attack operations in external storage, particularly by removing victim expansion and by removing squatting attacks. However, we find that many victims remain threatened because of the use of legacy apps (RQ3).   To assess the ideal impact of Scoped Storage, we evaluate the hypothetical case where all apps are compliant with Scoped Storage (RQ4).  We also examine two  vulnerabilities 
found in external storage using \tool and the performance of access control analysis using \tool in Sections~\ref{subsec:case} and~\ref{subsec:perf}, respectively.

\vspace{-0.1in}
\subsection{Effects of Scoped Storage}
\label{subsec:ess}
\vspace{-0.03in}

\textbf{RQ1:} {\em What fraction of the total number of threats in Android systems occur in external storage before and after the addition of Scoped Storage?} Table~\ref{tab:sum_table} displays the integrity violation (IVs) and attack operation (Attack Ops) counts for the eight Android systems, where two systems (Pixel3a 9.0 and Galaxy S20 9.0) do not use Scoped Storage.  Table~\ref{tab:sum_table} compares the total counts to the counts in external storage only.  Note that the IV counts ({\bf IVs}) are a sum of the number of objects that may be used to attack each victim, as described in Table~\ref{tab:sum_table}.  

The first two rows in Table~\ref{tab:sum_table} show the IV and attack operation counts for the whole system, showing that the number of attack operations tends to be slightly greater than the number of IVs for the system at large, although slightly lower for OnePlus systems.   The next two rows in Table~\ref{tab:sum_table} show the IV and attack operation counts for external storage alone.  For external storage, the number of attack operations is always less than the number of IVs, due to the lack symbolic links in external storage, which predates Scoped Storage, and the reduction in squatting attacks, which we show using \tool in Section~\ref{subsec:iv-ext}.  

Examining Table~\ref{tab:sum_table} from left to right, we see that the IV and attack operation counts of the pre-Scoped systems (i.e., Pixel3a 9.0 and Galaxy S20 9.0) were much higher than their respective counterparts (i.e., other Google devices and OEM devices, respectively).  \tool shows that this reduction is largely because Scoped Storage eliminates victim expansion (see Section~\ref{subsec:expand}), as described in Section~\ref{subsec:iv-ext}.  

However, we see the reduction in the fraction of IVs and attack operations differs between Google and OEM devices.  
For OEM devices, while the reduction in the number of IVs and attack operations is significant (i.e., from pre-Scoped Storage Galaxy S20 9.0 to Samsung and Oneplus versions 11.0 and 12.0), these counts remain much greater than for Google devices.  
We use \tool to assess how the greater use of legacy apps (i.e., apps not compliant with Scoped Storage) in these OEM devices increases these counts (see Section~\ref{subsec:legacy-app}) and the impact if all apps would be compliant with Scoped Storage (see Section~\ref{subsec:fss}).

\vspace{-0.1in}
\subsection{Reasons Scoped Storage Reduces Threats}
\label{subsec:iv-ext}
\vspace{-0.03in}

\begin{table*}[t]
\centering
\resizebox{\textwidth}{!}{
\begin{threeparttable}[b]
    \caption{Integrity Violations (IVs) by IV Type in Total and in External Storage across Vendor Releases}
    \label{tab:attack_surface}
     \footnotesize
\begin{tabular}{@{}c|ccc|ccccc@{}}
                                 & \multicolumn{3}{c|}{\small{\textbf{Google Devices}}} & \multicolumn{5}{c}{\small{\textbf{OEM Devices}}} \\
                                 &\textbf{{\em Pixel3a 9.0}} & \textbf{Pixel3a 11.0} & \textbf{Pixel3a 12.0} & \textbf{{\em Galaxy S20 9.0}}  & \textbf{Galaxy S20 11.0} & \textbf{Galaxy S20 12.0}  & \textbf{Oneplus8T 11.0} & \textbf{Oneplus8T 12.0} \\ \midrule
\textbf{Total File IVs (Read/Write)}           & 750/149  & 632/164               & 713/154   &13,248/7,213                      & 6,281/4,186            &3,036/1,797           &5,865/3,415        & 7,248/4,681  \\
\textbf{Total Dir IVs (Pathname/Binding)}        & 1,674/314  & 702/195                & 767/389  &18,241/6,713                      & 6,432/4,768            &3,772/2,667           &6,122/2,196        &7,420/2,687 \\ \hline
\textbf{Ext File IVs (Read/Write)}      &308/149 &202/132                   &187/111  &6,569/4,384   & 2,758/2,080 & 1,503/1,049 &2,944/977 & 3,879/1,293 \\
\textbf{Ext Dir IVs (Pathname/Binding)}   &713/219 &58/58 & 187/187 &6,110/4,767 & 955/955 & 785/785 & 1,421/1,421 & 1,653/1,653 \\ \hline
\end{tabular}%
\begin{tablenotes}\scriptsize
    \item[] {\bf IVs} $= \sum_v^V |IV_{obj}(v)|$, where $IV_{obj}(v)$ returns the set of objects in the IVs for a victim $v \in V$ as computed per Section~\ref{subsec:ivdef}.
    \item [] 
    Ext IVs are IVs whose objects are located in an external storage partition
\end{tablenotes}
\end{threeparttable}
}
\end{table*}

\textbf{RQ2:} {\em How does Scoped Storage impact the types of
integrity violations and attack operations that may be
attempted in external storage?}  In this section, we show how \tool enables us to explain the reasons for any reductions in threats due to Scoped Storage from Table~\ref{tab:sum_table}.  Rows 1-2 in Table~\ref{tab:attack_surface} show the total IV counts  for the eight Android systems in Table~\ref{tab:sum_table} broken down for file objects (read-IVs and write-IVs) and directories (pathname-IVs and binding-IVs).  Note that the write-IVs are a subset of the read-IVs and the binding-IVs are a subset of the pathname-IVs\footnote{The definition of binding-IVs implies that they are a subset of the pathname-IVs, but the write-IVs happen to be a subset of the read-IVs because victims always have read permission when they have write permission in the systems we examined.}  Rows 3-4 in Table~\ref{tab:attack_surface} show the same information, but for external storage only.



We can see in Table~\ref{tab:attack_surface} that the number of IVs for both files and directories are significantly reduced when compared to the pre-Scoped Storage systems (i.e., Pixel3a 9.0 and Galaxy S20 9.0).  This shows that access control decisions, such as the deprecation of the WEX
permission and more limited use of the 
REX permission reduce threats.  However, an even more obvious impact is shown in row 4, where the IV counts are the same for pathname-IVs and binding-IVs.  This is caused because adversaries cannot change the permissions of other apps, i.e., victim expansion as described in Section~\ref{subsec:expand} is no longer possible in Scoped Storage systems.

\begin{table*}[t]
\centering
\resizebox*{!}{3.5cm}{
\begin{threeparttable}%
    \caption{Attack Operations by Type in Total and in External Storage across Vendor Releases}
    \label{tab:ext_AOC}
     \footnotesize
\begin{tabular}{@{}c|ccc|ccccc@{}}
                                 & \multicolumn{3}{c|}{\small{\textbf{Google Devices}}} & \multicolumn{5}{c}{\small{\textbf{OEM Devices}}} \\
                                 & \textbf{{\em Pixel3a 9.0}} & \textbf{Pixel3a 11.0} & \textbf{Pixel3a 12.0}  & \textbf{{\em Galaxy S20 9.0}} &\textbf{Galaxy S20 11.0} & \textbf{Galaxy S20 12.0}  & \textbf{Oneplus8T 11.0} & \textbf{Oneplus8T 12.0} \\ \midrule
\textbf{Modification Attacks}              & 750 & 632                & 713                   & 13,248   &6,281 &3,036                & 5,865 & 7,284\\
\textbf{Squat Attacks}              & 314 & 195                & 389                   & 6,713   &4,768 &2,667                & 2,196 & 2,687\\ \hline
\textbf{Ext Modification Attacks}              & 308 & 202                & 187                   & 6,569   &2,758 &1,503                & 2,944 & 3,879\\
\textbf{Ext Squat Attacks}    & 219                  & 39                    & 32    & 4,767  & 461                 & 403 & 985 & 575\\
\textbf{Squat Attacks Prevented}          & 0                & 19                  & 155  & 0  & 454                  & 382 & 436 & 1,078\\ \hline
\end{tabular}%
\begin{tablenotes}\scriptsize
  \item[] {\bf Attack Ops} $= \sum_{iv}^{{\bf IVs}} |OP(iv)|$, where $OP(iv)$ returns the set of attack operations for an integrity violation $iv \in$ {\bf IVs}  computed per Section~\ref{subsec:operations}.
    \item[] Ext Attack Ops are Attack Ops whose objects are located in an external storage partition.
\end{tablenotes}
\end{threeparttable}%
}
\end{table*}

Table~\ref{tab:ext_AOC} shows the counts for the modification attack and squat attack operations described in Section~\ref{subsec:operations} in total and for external storage only\footnote{We do not show link traversal and luring traversal attack operations, which require symbolic links that have been banished from external storage since prior to Scoped Storage.}. 
Once again the counts for attack operations in external storage is significantly lower in Scoped Storage systems than pre-Scoped Storage systems (Pixel3a 9.0 and Galaxy S20 9.0).
Modification attacks are reduced because the number of file IVs has been reduced as discussed above. In addition, the number of squat attacks have been reduced because Scoped Storage prevents victims from accessing
adversary-created files by default (i.e., without REX or MES permission).
In row 5 of Table~\ref{tab:ext_AOC}, we show the count of the number of binding-IVs that cannot be converted into squat attacks because the victims lack REX or MES permissions to access adversary-controlled directories.  As a result, recent vulnerabilities, such as the Man-in-the-Disk~\cite{MITD} that leverage squat attack operations, are no longer possible by default  in Scoped Storage releases.    


\vspace{-0.13in}
\subsection{Problems with Legacy Applications}
\label{subsec:legacy-app}
\vspace{-0.02in}
\textbf{RQ3:} {\em How many of the threats in external storage are
due to legacy apps?} 
Table~\ref{tab:legacy} shows a comparison of IVs (i.e., broken down further into victim subject and object counts) created by apps that are compliant with Scoped Storage  and those that are not, which are called {\em legacy apps}.  Here, we see that a modest number of pre-installed legacy apps across vendors (row 1) causes over twice the number of subjects to become potential victims (i.e., have at least one IV due to a legacy app) of attacks (row 2) due to nearly twice the number of object (row 3) than for compliant apps (rows 4-6).  
This is not surprising since the access permissions of legacy apps 
is similar to Scoped Storage apps with MES Android permission. But, when applying Scoped Storage, the MES Android permission is only granted to applications that have been vetted by Google, which limits the number of third-party apps that may obtain that permission and presumably improves the trust in such apps.

\vspace{-0.13in}
\subsection{Fully-Enforced Scoped Storage}
\label{subsec:fss}
\vspace{-0.02in}
\textbf{RQ4:} {\em How many attack operations and adversaries could be removed if all legacy
apps are converted to Scoped Storage apps?} To measure how well Scoped Storage could  potentially work to reduce attack operations, we move objects in legacy locations into the shared folders protected by Scoped Storage (e.g., ownership info tracked by MediaProvider). Since the \emph{WRITE\_EXTERNAL\_STORAGE} permission has been deprecated beginning with Android 11, only file owners have write access to the files in shared folders. Then, we assume legacy flags are removed and perform \tool analysis to compute attack operations on external storage only. The analysis results are shown in Table~\ref{tab:full_ss}, where the row 1 is the original attack operation count for external storage and row 2 is the attack operation count for external storage after the procedures described above, and rows 3 and row 4 count the corresponding changes in the number of adversaries. 

We see that the number of attack operations decreases 12\%-28\%, but the number of adversaries decreases more significantly: at least 36\% for Google devices and at least 65\% for all other OEM devices.  This suggests that 
the level of decrease in attack operations does not reflect the corresponding reduction in the number of adversaries. We observe that the remaining adversaries are file management apps given MES and REX permissions in Scoped Storage, which conflicts with their low privilege classification under the Google Privilege Levels~\cite{PPRIV_LEVEL} in Table~\ref{table:levels}. It is future work to assess whether permissions should be refined further to reduce attack operations or additional privilege levels need to be added to accommodate such apps. 

\vspace{-0.12in}
\subsection{Vulnerability Case Studies}
\label{subsec:case}
\vspace{-0.03in}

Using the attack operations computed by \tool, we manually identified two previously unknown vulnerabilities that we describe below as well as other resources that face significant risks.  We have ethically reported these vulnerabilities.

\paragraph{Replace Over-the-Air Updates}
We found a new vulnerability in the Oneplus 8T system running Android 11 release. We found that Oneplus temporarily stores an OTA update  file in a hidden folder located in the root directory of external storage (i.e., accessible to legacy apps). We found that untrusted applications that request the legacy  apps can observe the OTA download and replace the OTA update file before installation takes place. This could potentially grant root privilege to attackers with a properly engineered OTA update file. \tool further identified many other objects stored in legacy location vulnerable to adversarial legacy apps. These files include configuration files, log files, cache files, and cookies. We did not fully explore how these attack operations can be exploited, but it is extremely dangerous for privileged apps to store their data files in locations accessible to legacy apps. The above vulnerability shows the danger of legacy apps, and how OEMs may not use external storage correctly in the face of legacy apps.

\paragraph{Malicious Code Execution}
We found potential vulnerability related to the Quick App feature that is widely used by major Chinese OEMs including Huawei, Lenovo, Oneplus and Xiaomi. Quick App is a lightweight framework that allows users to access simple services (i.e., weather, taxi, payment) without installing heavyweight APKs. One of the most used features is gaming, where users can start playing with one simple click. On the Oneplus 8T Android 11 device we tested, \tool found that the Quick App framework stores executable files in a hidden folder located in a legacy location, where malicious legacy apps can easily squat. The Quick App framework is a highly privileged victim running as a pre-installed platform app. We did not fully evaluate how much damage we can done to the system, but we are able to cause the Quick App service to restart by corrupting the game files.

\begin{table*}[t]
\centering
\resizebox*{!}{3.5cm}{
\begin{threeparttable}%
    \caption{Number of Apps and Their  Victims (for Attack Operations) for Scoped-Apps vs. Legacy-Apps}
    \label{tab:legacy}
     \footnotesize
\begin{tabular}{@{}c|cc|cccc@{}}
                                 & \multicolumn{2}{c|}{\textbf{Google Devices}} & \multicolumn{4}{c}{\textbf{OEM Devices}} \\
                                 & \textbf{Pixel3a 11.0} & \textbf{Pixel3a 12.0}  & \textbf{Galaxy S20 11.0} & \textbf{Galaxy S20 12.0}  & \textbf{Oneplus8T 11.0} & \textbf{Oneplus8T 12.0} \\ \midrule
\textbf{Legacy-App Count}    & 11                  & 18                    & 44    &44  &23                 & 22\\
\textbf{Victims of Legacy-Apps$^1$}        & 141                 & 254                   & 280   &286 &275            & 293\\ 
\textbf{Object Count} &6 & 4 & 44 & 24 & 53 & 57\\ \hline
\textbf{Scoped-App Count}              & 106                & 252                   & 230   &205 &222                & 215\\
\textbf{Victims of Scoped-Apps$^2$}          & 69                & 124                  & 120  &125  &113                  & 108\\
\textbf{Object Count} &4 & 2 & 12 & 10 & 31 & 33\\
\end{tabular}%
\begin{tablenotes}\scriptsize
    \item [] Unit: Subject Count
    \item[1] Number of unique victims with IVs where an adversary is a legacy app.
    \item[2] Numbers of unique victims with IVs where an adversary is a compliant app
\end{tablenotes}
\end{threeparttable}%
}
\end{table*}

\begin{table*}[t]
\centering
\resizebox*{!}{3.1cm}{
\begin{threeparttable}%
    \caption{Attack Operation Comparison between Current Systems and Fully Enforced Scoped Storage}
    \label{tab:full_ss}
     \footnotesize
\begin{tabular}{@{}c|cc|cccc@{}}
                                 & \multicolumn{2}{c|}{\textbf{Google Devices}} & \multicolumn{4}{c}{\textbf{OEM Devices}} \\
                                 & \textbf{Pixel3a 11.0} & \textbf{Pixel3a 12.0}  & \textbf{Galaxy S20 11.0} & \textbf{Galaxy S20 12.0}  & \textbf{Oneplus8T 11.0} & \textbf{Oneplus8T 12.0} \\ \midrule
\textbf{Ext-Storage Attack Operations}    & 241                  & 219                    & 3,219    & 1,906                 & 3,929 & 4,454\\
\textbf{Full-Scoped Attack Operations$^1$}              & 173(-28\%)                & 166(-24\%)                   & 2,831(-12\%)   &1,620(-15\%) & 3,222(-18\%)        & 3,564(-20\%)\\ \hline
\textbf{Ext-Storage Adversaries}        & 25        & 22          & 62   &57  &61            & 63\\
\textbf{Full-Scoped Adversaries$^2$}    & 16 (-36\%)       & 9(-59\%)           & 18(-70\%)   &18(-68\%)  &21(-65\%)            & 16(-74\%)\\
\end{tabular}%
\begin{tablenotes}\scriptsize
    \item[1] Fully enforced Scoped Storage attack operation count
    \item[2] Numbers of unique attackers after Scoped Storage is fully enforced
\end{tablenotes}
\end{threeparttable}%
}
\end{table*}


\vspace{-0.12in}
\subsection{\tool Analysis Performance}
\label{subsec:perf}
\begin{figure}[t]
\centering
\includegraphics[width=8cm]{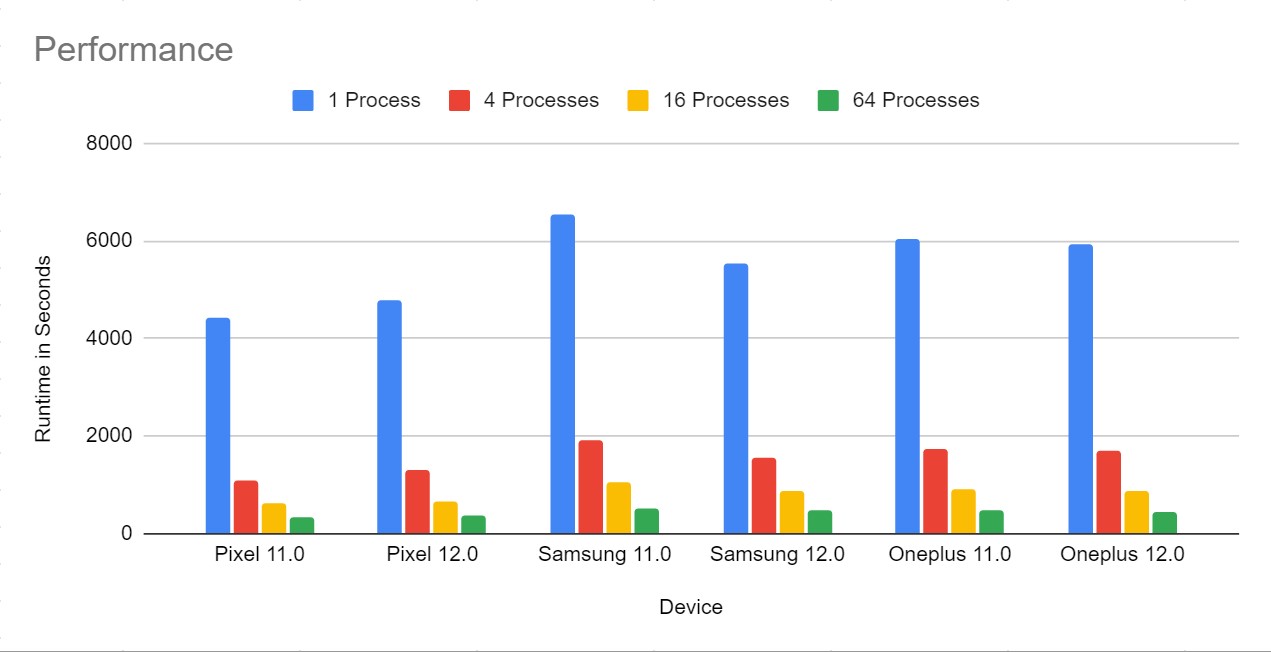}
\vspace{-1.5em}
\caption{{\tool Analysis Performance}}
\vspace{-0.12in}
\label{fig:perf}
\end{figure}

\vspace{-0.05in}
We measured the performance of \tool in analyzing the six Android releases supporting Scoped Storage.  The overhead was measured on a Mac M1 Pro (10 cores) with 16GB of RAM.  We measure the performance of the \tool Main and Workers to compute attack operations as described in Section~\ref{sec:impl}.  
Unlike the original \tool tool~\cite{lee21usenix}, we use a multi-process implementation to leverage parallelism for \tool Workers, as described in Section~\ref{sec:impl}.  Initially, we divided the objects among the workers evenly, but we found that this results in an unbalanced load as some objects have many more readers and writers than others.  As a result, we assign the objects one at a time to workers, which tends to balance the load.  
Figure~\ref{fig:perf} shows the performance results.  We measure the performance of \tool's analysis using 1 to 64 processes. We can see that performance benefits significantly as we increase the number of processes, from over 6,000s for a single process to a maximum of 524s for 64 processes for the six Android systems.

\vspace{-0.12in}
\section{Discussion}
\label{sec:discussion}
\vspace{-0.05in}



\paragraph{Limitations of \tool} We identify three limitations of the current \tool tool: (1) \tool requires a rooted phone to collect filesystem data; (2) \tool cannot always determine the mapping among multiple access control policies for all subjects; (3) \tool cannot confirm vulnerabilities from attack operations automatically.  

Without rooting the phone, we cannot gather DAC information from privileged directories such as {\tt /system}.  Recently,  Hernandez et al.~\cite{BigMAC} proposed BigMAC, which includes a technique to extract accurate DAC configuration data from these privileged directories ($\sim$95\%).  However, we found that BigMAC cannot extract files effectively from some directories, such as {\tt /data}, as described under Data Collection in Section~\ref{sec:impl}. 
We will explore methods to achieve complete recovery in future work.

Another limitation of \tool is that we cannot map all processes to complete subjects as defined in Section~\ref{subsec:label}.  The main problem is to connect the package name and UID from the package list (see Section~\ref{sec:impl} for \tool Main) to the MAC label/MLS category.  Recall that the mapping between UID and MAC label is determined by running the program~\cite{chen17acsac}, but not all packages listed can be run, as some have abstract names.  We compare the number of entries in the package list to the number of unique subjects we compute per system.  For Google devices, the count is the same, but for OEM devices up to 10\% of the package list entries are unmapped (for Samsung Galaxy 11.0 and 12.0).  We will explore how to complete this mapping in the future.

Finally, \tool lacks a systematic way to test the victims for vulnerabilities to the attack operations found.  There are two problems to address.  First, we need to know when a victim may use a resource that is associated with an attack operation.  This is difficult to determine statically.  The STING system~\cite{sting} provides passive runtime monitoring of processes for use of bindings that could be used to perform file squatting and link traversal attacks using DAC policies, so such a runtime monitoring approach could be extended to utilize attack operations generated by \tool. 
Second, once we know when a victim may be threatened by an attack operation, we need to generate test cases that could exploit the victim.  Current fuzzing techniques~\cite{chizpurfle} do not target these types of attack operations.  Runtime monitoring techniques~\cite{sting,jigsaw14usenix} and similar techniques for assessing use of Android intents~\cite{aafer15ccs} generate simple test cases, enabling detection of unprotected cases.  We aim to generate test cases that account for the conditional checks in the program fully. 

\paragraph{Limitation of Scoped Storage}
In terms of Scoped Storage, most of the security problems are caused by OEMs not following the safe guidance or using the legacy flag as shown in Section~\ref{subsec:legacy-app}. We expect the problem to ease and disappear when Scoped Storage is fully enforced. However, attack operations caused by third party application with MES shown in Section~\ref{subsec:fss} will still exist and we believe that a more fine-grained permission control is needed. Potential methods include new data access API specifically for MES apps or new resources protection technique for all applications. An intermediate solution will be limiting MES permission usage while apps are in the background, or notify users when apps are using MES.

\vspace{-0.1in}
\section{Related Work}
\label{sec:relwork}
\vspace{-0.07in}


Researchers have long known about the three types of integrity violations listed in Section~\ref{sec:threat}, but have found it difficult to vulnerabilities to such violations.  A variety of mechanisms have been proposed to prevent attacks during name resolution, including defenses for binding and pathname vulnerabilities.  These defenses have often been focused on TOCTTOU attacks~\cite{mcphee74,bishop-dilger, ExtRc}.  Some defenses are implemented in the program or as library extensions~\cite{raceguard,27,dean-hu,tsafir} and some as kernel extensions~\cite{chapin_tocttou,28,venema_ndss_2010,openwall,ty-race,35}, but the methods overlap, where some enforce invariants on file access~\cite{raceguard,chapin_tocttou,35,27,28,ty-race}, some enforce namespace invariants~\cite{venema_ndss_2010,openwall}, and some aim for ``safe'' access methods~\cite{dean-hu,tsafir}.  In general, all program defenses have been limited because they lack insight into the system state and all system defenses are limited because they lack side-information about the intent of the program~\cite{johnson-tocttou}.  


The main defense for preventing filesystem vulnerabilities is access control.  If the access control policies prevent an adversary from accessing the filesystem resources that enable attack operations, then the system is free of associated vulnerabilities.  However, the discretionary access control (DAC) policies commonly used do not enable prediction of whether a subject may obtain an unauthorized permission~\cite{hru76}, so techniques to restrict DAC~\cite{take-grant,tam,tidswell00ccs} and apply mandatory access control (MAC) enforcement~\cite{blp76,b77} were then explored, culminating in MAC enforcement systems, such as Linux Security Modules~\cite{wcs+02} employed by SELinux~\cite{selinux} and AppArmor~\cite{apparmor}.  Researchers than proposed MAC enforcement for Android systems~\cite{xie09srds,bugiel12ndss}, so a version of SELinux~\cite{selinux} targeting Android was developed, called Security Enhanced (SE) Android~\cite{smalley2013security}.  However, the attack operations we find in this paper abuse available MAC enforcement.  While a techniques have been developed to limit the permissions available to individual system calls~\cite{cw-lite,process-firewall}, such techniques need policy analysis to determine the policies to enforce.



Researchers have proposed using access control policy analysis to identify misconfigurations that may lead to vulnerabilities~\cite{jaeger02sacmat,setools}, but traditionally, access control policy analysis methods only reason about one policy, such as a mandatory access control (MAC) policy~\cite{setools,jaeger03usenix,chen09ndss,integrity-wall} or an Android permission policy~\cite{enck09ccs,wae+17,Wang2015}. However, based on the research challenges above, we must consider the combination of the access control policies employed on the system to compute attack operations accurately.  Chen {\em et al.}~\cite{chen17acsac} were the first work that we are aware of to combine MAC and DAC policies in access control policy analysis. Hernandez {et al.}~\cite{BigMAC} further extended their analysis to include MAC, DAC and Linux capabilities.  However, both of these techniques compute data flows, which are much more numerous than integrity violations.

\vspace{-0.1in}
\section{Conclusions}
\label{sec:conc}
\vspace{-0.05in}
Android uses a combination of filesystem access control mechanisms to assure its platform integrity.
This paper has proposed \tool, a policy analysis tool that reasons over Android's mandatory (SELinux)~\cite{SEAndroid_G} and discretionary (Unix permissions) access control policies, in addition to the other mechanisms (e.g., Android permissions, Scoped Storage) that influence file access control.
\tool is novel in its ability to reason about permission expansion, which lies at the intersection of mandatory and discretionary policy.
We applied \tool to three different Google Android releases and five different OEM Android releases, characterizing the potential for file-based attacks such as file squatting, link traversal, and luring traversal. We perform quantitative evaluation on the new Scoped Storage technique highlighting its benefit and current problems. Our results suggest that Scoped Storage is indeed effective and helps with reducing the attack vector of Android systems. However, legacy applications could still produce problems and OEMs must make privileged applications Scoped Storage compliant as soon as possible.

\begin{spacing}{0.95}
\printbibliography

@INPROCEEDINGS{wae+17,
  AUTHOR = {Ruowen Wang and Ahmed M. Azab and William Enck and Ninghui Li and Peng Ning and Xun Chen and Wenbo Shen and Yueqiang Cheng},
  TITLE = {{SPOKE: Scalable Knowledge Collection and Attack Surface Analysis of Access Control Policy for Security Enhanced Android}},
  BOOKTITLE = {Proceedings of the ACM Asia Conference on Computer and Communications Security (ASIACCS)},
  YEAR = {2017},
}

@misc{rahman_2022, title={How scoped storage changed Android File Access - Android bytes podcast}, url={https://blog.esper.io/how-scoped-storage-works/}, journal={Esper Blog}, author={Rahman, Mishaal}, year={2022}, month={Oct}}

@INPROCEEDINGS{chen17acsac,
  AUTHOR = {Haining Chen and Ninghui Li and William Enck and Yousra Aafer and Xiangyu Zhang},
  TITLE = {{Analysis of SEAndroid Policies: Combining MAC and DAC in Android}},
  BOOKTITLE = {Proceedings of the Annual Computer Security Applications Conference (ACSAC)},
  YEAR = {2017},
}

@inproceedings{smalley2013security,
  title={{Security Enhanced (SE) Android: Bringing Flexible MAC to Android.}},
  author={Smalley, Stephen and Craig, Robert},
  booktitle={Proceedings of the 20th Network and Distributed Systems Symposium (NDSS)},
  year={2013}
}

@online{MITD,
  title={{Man-in-the-Disk:Android Apps Exposed via External Storage}}, 
  url = {https://research.checkpoint.com/2018/androids-man-in-the-disk/},
  journal={Check Point Research}, 
  author={Makkaveev,Slava}, 
  year={2019}, 
  month={Feb}
}

@online{cve202220220,  
  title={{Android Security Bulletin-July 2022}}, 
  Howpublished={\url{https://source.android.com/docs/security/bulletin/2022-07-01/}}, 
  author={Google}, 
  year={2022}, 
  month={July}
}

@misc{SEAndroid_G, title={{Security-enhanced Linux in Android}}, url={https://source.android.com/docs/security/features/selinux}, journal={Android Open Source Project}, publisher={Google}, year={2022}, month={Sep}}

@inproceedings {tfntf,
author = {Bharath Kumar Reddy Vangoor and Vasily Tarasov and Erez Zadok},
title = {To {FUSE} or Not to {FUSE}: Performance of {User-Space} File Systems},
booktitle = {15th USENIX Conference on File and Storage Technologies (FAST 17)},
year = {2017},
isbn = {978-1-931971-36-2},
address = {Santa Clara, CA},
pages = {59--72},
url = {https://www.usenix.org/conference/fast17/technical-sessions/presentation/vangoor},
publisher = {USENIX Association},
month = feb,
}

@online{PPRIV_LEVEL,
  title = {{Security Updates and Resources}},
  author = {Google},
  url = {https://source.android.com/docs/security/overview/updates-resources#process_types},
  note = {Accessed Dec. 10, 2022.},
  YEAR = {2022}
}

@online{setools,
  title = {{SETools}},
  url = {https://github.com/TresysTechnology/setools},
  note = {Accessed Dec 2022},
}

@inproceedings{Wang2015,
 author = {Wang, Ruowen and Enck, William and Reeves, Douglas and Zhang, Xinwen and Ning, Peng and Xu, Dingbang and Zhou, Wu and Azab, Ahmed M.},
 title = {{EASEAndroid: Automatic Policy Analysis and Refinement for Security Enhanced Android via Large-scale Semi-supervised Learning}},
 booktitle = {Proceedings of the 24th USENIX Conference on Security Symposium},
 year = {2015},
 pages = {351--366},
}

@online{statcounter,
  AUTHOR = {{StatCounter}},
  TITLE = {{OS Market Share}},
  MONTH = mar,
  YEAR = 2020,
  url =
  {https://gs.statcounter.com/os-market-share},
}

@inproceedings{cw-lite,
  author    = {Umesh Shankar and
               Trent Jaeger and
               Reiner Sailer},
  title     = {{Toward Automated Information-Flow Integrity Verification for Security-Critical
               Applications}},
  booktitle = {Proceedings of the 2006 Network and Distributed System Security Symposium
               (NDSS)},
  year      = {2006},
}

@inproceedings{xie09srds,
  author    = {Liang Xie and
               Xinwen Zhang and
               Ashwin Chaugule and
               Trent Jaeger and
               Sencun Zhu},
  title     = {{Designing System-Level Defenses against Cellphone Malware}},
  booktitle = {28th {IEEE} Symposium on Reliable Distributed Systems (SRDS)},
  year      = {2009},
}

@inproceedings{tidswell00ccs,
  author    = {Jonathon Tidswell and
               Trent Jaeger},
  title     = {{An access control model for simplifying constraint expression}},
  booktitle = {Proceedings of the 7th {ACM} Conference on Computer and
               Communications Security},
  year      = {2000},
}

@InProceedings{jaeger03usenix,
  author =       "Trent Jaeger and Reiner Sailer and Xiaolan Zhang",
  title =	{{Analyzing Integrity Protection in the SELinux
		 Example Policy}},
  booktitle =    "Proceedings of the $12^{th}$ {USENIX} Security Symposium",
  year =	 "2003",
}

@inproceedings{chen09ndss,
  author={Chen, Hong and Li, Ninghui and Mao, Ziqing},
  title= {{Analyzing and Comparing the Protection Quality of Security Enhanced Operating Systems}},
  booktitle={Proceedings of the 16th Network and Distributed System Security Symposium(NDSS)},
  pages={11--16},
  year={2009}
}

@inproceedings{enck09ccs,
author = {Enck, William and Ongtang, Machigar and McDaniel, Patrick},
title = {{On Lightweight Mobile Phone Application Certification}},
year = {2009},
booktitle = {Proceedings of the 16th ACM Conference on Computer and Communications Security},
pages = {235–245},
}

@inproceedings{jaeger02sacmat,
author = {Jaeger, Trent and Edwards, Antony and Zhang, Xiaolan},
title = {{Managing Access Control Policies Using Access Control Spaces}},
year = {2002},
address = {New York, NY, USA},
booktitle = {Proceedings of the Seventh ACM Symposium on Access Control Models and Technologies},
pages = {3–12},
}

@inproceedings{aafer15ccs,
author = {Aafer, Yousra and Zhang, Nan and Zhang, Zhongwen and Zhang, Xiao and Chen, Kai and Wang, XiaoFeng and Zhou, Xiaoyong and Du, Wenliang and Grace, Michael},
title = {{Hare Hunting in the Wild Android: A Study on the Threat of Hanging Attribute References}},
year = {2015},
booktitle = {Proceedings of the 22nd ACM Conference on Computer and Communications Security},
pages = {1248–1259},
}

@inproceedings{integrity-wall,
author = {Vijayakumar, Hayawardh and Jakka, Guruprasad and Rueda, Sandra and Schiffman, Joshua and Jaeger, Trent},
title = {{Integrity Walls: Finding Attack Surfaces from Mandatory Access Control Policies}},
year = {2012},
booktitle = {Proceedings of the 7th ACM Symposium on Information, Computer and Communications Security},
pages = {75–76},
}

@inproceedings{jigsaw14usenix,
  author    = {Hayawardh Vijayakumar and Xinyang Ge and Mathias Payer and
               Trent Jaeger},
  title     = {{Jigsaw: Protecting Resource Access by Inferring Programmer Expectations}},
  booktitle = {Proceedings of the $23^{rd}$ USENIX Security Symposium},
  year      = {2014},
  month     = AUG,
}

@inproceedings{process-firewall,
  author    = {Hayawardh Vijayakumar and Joshua Schiffman and Trent Jaeger},
  title     = {{Process Firewall: Protecting Processes During Resource Access}},
  booktitle = {Proceedings of the Eighth European Conference on Computer Systems},
  year      = {2013}
}

@article{mcphee74,
 author = {McPhee, W. S.}, 
 title = {{Operating System Integrity in OS/VS2}},
 journal = {IBM System Journal}, 
 volume = {13},
 issue = {3}, 
 month = {September},
 year = {1974},
 pages = {230--252},
 numpages = {23},
 publisher = {IBM Corp.},
}

@article{hru76,
	Author = {Michael Harrison and Walter Ruzzo,  and Jeffrey Ullman},
	Journal = {Communications of ACM},
	Month = aug,
	Title = {{Protection in Operating Systems}},
	Year = 1976}

@inproceedings{flask-inevitability,
	Author = {Peter {Loscocco \emph{et al.}}},
	Booktitle = {Proceedings of the 21st National Information Systems Security Conference},
	Pages = {303--314},
	Title = {{The Inevitability of Failure: The Flawed Assumption of Security in Modern Computing Environments}},
	Year = {1998}}

@inproceedings{take-grant,
	Author = { Richard Lipton and Lawrence Snyder},
	Booktitle = {Proceedings of the $17^{th}$ Annual Symposium on Foundations of Computer Science},
	Title = {{A Linear Time Algorithm for Deciding Security}},
	Year = {1976}}

@inproceedings{tam,
	Author = {Ravi Sandhu},
	Booktitle = {Proceedings of the 1992 IEEE Symposium on Security and Privacy},
	Title = {{The Typed Access Matrix Model}},
	Year = {1992}}

@article{lee21ieeesp,
    Author = {{Yu-Tsung} Lee and Haining Chen and Trent Jaeger},
	Journal = {IEEE Security \& Privacy},
	Volume = 19,
	Number = 5,
	Title = {{Demystifying Android's Scoped Storage Defense}},
	Year = 2021
}

@inproceedings{lee21usenix,
   author = {{Yu-Tsung} Lee and William Enck and Haining Chen and Hayawardh
	Vijayakumar and Ninghui Li and Daimeng Wang and Zhiyun Qian and
	Giuseppe Petracca and Trent Jaeger},
   title = {{PolyScope: Multi-policy} Access Control Analysis to Compute Authorized Attack Operations in {Android} Systems},
   booktitle = {Proceedings of the 30th USENIX Security Symposium},
   month = aug,
   year = {2021}
}

@inproceedings{bugiel12ndss,
  author    = {Sven Bugiel and
               Lucas Davi and
               Alexandra Dmitrienko and
               Thomas Fischer and
               Ahmad{-}Reza Sadeghi and
               Bhargava Shastry},
  title     = {{Towards Taming Privilege-Escalation Attacks on Android}},
  booktitle = {Proceedings of the 19th Network and Distributed System Security Symposium(NDSS)},
  year      = {2012},
}

@article{confused-deputy,
	Author = {Norm Hardy},
	Issn = {0163-5980},
	Journal = {ACM Special Interest Group in Operating Systems, Operation System Review},
	Number = {4},
	Publisher = {ACM},
	Title = {{The Confused Deputy: or Why Capabilities Might Have Been Invented}},
	Volume = {22},
	Year = {1988}}

@online{selinux,
  title = {{SELinux}},
  url ={https://github.com/SELinuxProject},
  note = {(Accessed Dec 2022)},
  year = {-}
}

@inproceedings {BigMAC,
author={Grant Hernandez and Dave Jing Tian and Yadav, Anurag Swarnim and Williams, Byron J and Butler, Kevin RB},
title = {{BigMAC: Fine-Grained Policy Analysis of Android Firmware}},
booktitle = {Proceedings of the USENIX Security Symposium},
year = {2020},
}

@inproceedings{chizpurfle,
  title={{Chizpurfle: A Gray-box {Android} Fuzzer for Vendor Service Customizations}},
  author={Iannillo, Antonio Ken and Natella, Roberto and Cotroneo, Domenico and Nita-Rotaru, Cristina},
  booktitle={Software Reliability Engineering (ISSRE), IEEE 28th International Symposium},
  pages={1--11},
  year={2017},
}

@online{scopedstorage,
    title= {{Storage Updates in Android 11}},
    url = {https://developer.android.com/preview/privacy/storage},
    author = {Google},
    note = {Accessed June 2022}
}

@inproceedings{ExtRc,
author = {Du, Shaoyong and Liu, Xin and Lai, Guoqing and Luo, Xiangyang},
title = {Watch Out for Race Condition Attacks When Using Android External Storage},
year = {2022},
url = {https://doi.org/10.1145/3548606.3560666},
booktitle = {Proceedings of the 2022 ACM SIGSAC Conference on Computer and Communications Security},
pages = {891–904},
numpages = {14},
location = {Los Angeles, CA, USA},
series = {CCS '22}
}

@inproceedings{28,
  author    = {Calton Pu and Jinpeng Wei},
  title     = {{Modeling and Preventing TOCTTOU Vulnerabilities in Unix-style
               Filesystems}},
  booktitle   = {{IEEE International Symposium of System Engineering}},
  year      = {2006}}

@INPROCEEDINGS{35,
AUTHOR = "Prem Uppuluri and Uday Joshi and Arnab Ray",
TITLE = {{Preventing Race Condition Attacks on Filesystems.}},
booktitle = "ACM Symposium on Applied Computing",
YEAR = {2005}  }

@INPROCEEDINGS{1,
AUTHOR = "Ashish Aggarwal and Pankaj Jalote",
TITLE = "Monitoring the Security Health of Software Systems.",
booktitle = "ISSRE-06",
PAGES = {146-158},
YEAR = {2006}  }

@INPROCEEDINGS{27,
author={Park, Jongwoon and Lee, Gunhee and Lee, Sangha and Kim, Dong-kyoo},
TITLE = {{RPS: An Extension of Reference Monitor to Prevent Race-Attacks}},
booktitle = {Advances in Multimedia Information Processing},
YEAR = {2004}  }

@incollection{jaeger_refmon,
  author    = {Trent Jaeger},
  editor    = {Henk C. A. van Tilborg and
               Sushil Jajodia},
  title     = {Reference Monitor},
  booktitle = {Encyclopedia of Cryptography and Security, 2nd Ed},
  pages     = {1038--1040},
  publisher = {Springer},
  year      = {2011},
}

@article{chapin_tocttou,
  author    = {Kyung-suk Lee and
               Steve J. Chapin},
  title     = {{Detection of File-based Race Conditions}},
  journal   = {International Journal of Information Security},
  year      = {2005},
}

@inproceedings{tsafir,
	Author = {Tsafrir, Dan and Hertz, Tomer and Wagner, David and
Da Silva, Dilma},
	Booktitle = {USENIX Conference on File and Storage Technologies},
	Title = {{Portably Solving File TOCTTOU Races with Hardness Amplification}},
	Year = {2008},
}

@inproceedings{johnson-tocttou,
	Author = {Xiang Cai and Yuwei Gui and Rob Johnson},
	Booktitle = {{IEEE} Statistical Signal Processing Workshop},
	Title = {{Exploiting Unix File-System Races via Algorithmic Complexity Attacks}},
	Year = {2009},
}

@inproceedings{dean-hu,
	Author = {Drew Dean and Alan Hu},
	Booktitle = {Proceedings of the 13th conference on USENIX Security Symposium},
	Title = {{Fixing Races for Fun and Profit}},
	Year = {2004}}

@article{bishop-dilger,
	Author = {Matt Bishop and Michael Dilger},
	Journal = {Computer Systems},
	Month = {Spring},
	Number = 2,
	Title = {Checking for Race Conditions in File Accesses},
	Volume = 9,
	Year = 1996}

@inproceedings{jsz03,
	Author = {Jaeger, Trent and Sailer, Reiner and Zhang, Xiaolan},
	Booktitle = {Proceedings of the 12th USENIX Security Symp.},
	Month = aug,
	Title = {Analyzing Integrity Protection in the {SELinux} Example Policy},
	Year = 2003}

@inproceedings{wcs+02,
	Author = {Chris Wright and Crispin Cowan and James Morris},
	Booktitle = {USENIX Security Symposium},
	Title = {{Linux Security Modules: General Security Support for the Linux Kernel}},
	Year = 2002}

@techreport{b77,
	Author = {Kenneth Biba},
	Institution = {MITRE},
	Month = {April},
	Number = {MTR-3153},
	Title = {{Integrity Considerations for Secure Computer Systems}},
	Year = 1977}

@techreport{blp76,
	Author = {Bell Elliott, and Leonard La Padula},
	Institution = {Deputy for Command and Management Systems, HQ Electronic Systems Division (AFSC)},
	Month = {March},
	Number = {ESD-TR-75-306},
	Title = {{Secure Computer System: {Unified} Exposition and {Multics} Interpretation}},
	Year = 1976}

@inproceedings{raceguard,
 author = {Crispin Cowan and Steve Beattie and Chris Wright and Greg
Kroah-hartman},
 title = {{RaceGuard: Kernel Protection from Temporary File Race Vulnerabilities}},
 booktitle = {Proceedings of the 10th conference on USENIX Security Symposium},
 year = {2001},
}

@inproceedings{ty-race,
    author = {Eugene Tsyrklevich and Bennet Yee},
    title = {{Dynamic Detection and Prevention of Race Conditions in File Accesses}},
    booktitle = {USENIX Security Symposium}, 
    year = {2003},
}

@misc{apparmor,
	Author = {Novell},
	Howpublished = {\url{http://www.novell.com/linux/security/apparmor/}},
	Title = {{AppArmor Linux Application Security}},
}

@inproceedings{venema_ndss_2010,
  author    = {Suresh Chari and Shai Halevi and Wietse Venema},
  title     = {{Where Do You Want to Go Today? Escalating Privileges by
               Pathname Manipulation}},
  booktitle = {Proceedings of the 17th Network and Distributed System Security Symposium(NDSS)},
  year      = {2010}
}

@misc{openwall,
	Howpublished = {\url{http://www.openwall.com/}},
	Key = {openwall},
	Title = {{OpenWall Project - Information Security Software for Open Environments}},
	Year = {2008}}

@inproceedings{sting,
    author = {Hayawardh Vijayakumar and Joshua Schiffman and Trent
Jaeger},
    title = {{STING: Finding Name Resolution Vulnerabilities in Programs}},
    booktitle = {21st {USENIX} Security Symposium},
    year = {2012},
}
\end{spacing}

\end{document}